\documentclass[preprint,11pt]{JHEP3} 

\JHEPspecialurl{http://jhep.sissa.it/JOURNAL/JHEP3.tar.gz}

\usepackage{epsfig,multicol}

\usepackage{epsfig}
\usepackage{graphicx}
\usepackage{dcolumn}
\usepackage{bm}
\usepackage[hang,bf,small]{caption}
\usepackage{amsmath}
\usepackage{rotating}
\DeclareGraphicsExtensions{.eps.gz,.eps,.ps,.ps.gz}
\parskip=\itemsep

 \newcommand{\SP}{\langle \mid S \mid^2 \rangle}
\newcommand{\beq}{\begin{equation}}
\newcommand{\eeq}{\end{equation}}
\newcommand{\be}{\begin{eqnarray}}
\newcommand{\ee}{\end{eqnarray}}

\newcommand{\as}{\alpha_S}
\newcommand{\Lb}{\left(}
\newcommand{\Rb}{\right)}

\newcommand{\h}{\textstyle\frac{1}{2}}

\def\eq#1{{Eq.~(\ref{#1})}}
\def\fig#1{{Fig.~\ref{#1}}}
\def\gev{{\,\mbox{GeV}  }}
\def\gevs{{\,\mbox{GeV}^2}}

\def \gevms{{\,\mbox{GeV}^{-2}}}
\relax
\title{\huge \bf  Survival probability for exclusive central diffractive production of colorless states at the LHC.}
\author{\Large \bf E.~Gotsman ${}^{a)}$ \thanks{Email: gotsman@post.tau.ac.il}\,\,,\,\,
H.~Kowalski ${}^{b)}$\thanks{Email: henri.kowalski@desy.de}
\,\,,\,\,
E.
Levin ${}^{a)}$\thanks{Email:
leving@post.tau.ac.il, levin@mail.desy.de;}\,\, \,\,,\,\, U.~Maor ${}^{a)}$ \thanks{Email:
maor@post.tau.ac.il}\,\, \, and\,\, \, A.
Prygarin ${}^{a)}$\thanks{Email:
prygarin@post.tau.ac.il;} \\
 $a)$
Department of Particle Physics, School of Physics and Astronomy\\ Raymond and Beverly Sackler Faculty
of Exact Science\\  Tel Aviv University, Tel Aviv, 69978, Israel\\
$b)$ DESY,  ZEUS collaboration, 22603, Hamburg, Germany\\
}

\abstract{In this paper we discuss the survival probability for
exclusive central diffractive production of a colorless small size
system at the LHC. This process has a clear signature of two large
rapidity gaps. Using the eikonal approach for the description of
soft interactions, we predict the value of the survival probability
to be about  $ 5 \sim 6 \%$ for single channel models, while for a
two channel model the survival probability is about $3\%$. The
dependence of the survival probability factor (damping factor) on
the transverse momenta of the recoiled protons is discussed, and we
suggest it  be measured at the Tevatron so as to minimize the
possible ambiguity in the calculation of survival probability at the
LHC. }


\preprint{ \bf
{\tt  TAUP-2816-05} \\
{\tt \today}\\
{\tt hep-ph/0512254}
}

\begin{document}

\setcounter{page}{1}
\def\thefootnote{\arabic{footnote}}

\section{Introduction}
\setcounter{equation}{0}
Following the pioneering papers of Ref.\cite{DOK},
it has been realized\cite{BJ,GLM} that large rapidity gaps (LRG)
diffractive processes are suppressed due to the rescattering of the
spectator partons. The calculations of the resulting gap survival
probability (SP)\cite{GLMLRG,GLMSP,CZ,FLST,DG,DGK,BLHL}
have  since then been at the focus of high energy
phenomenology. The incentive
for this activity has been the need to obtain reliable SP
estimates for Higgs diffractive production at the LHC.
The analogous process of diffractive hard di-jet production has
been measured at Fermilab\cite{Fermi} and HERA\cite{HERA} and  this
may serve as a
laboratory to check the validity and reliability of the proposed
calculations.

SP is the probability to have a simple diffractive colorless final
state configuration, in which the LRG are preserved, regardless of
the strong interaction of the rescattered soft partons (see
\fig{lrg}). Its calculation depends crucially on soft scattering
physics, for which a theory is still lacking. The main
phenomenological tool utilized in these calculations is the Regge
soft Pomeron. In spite of the progress made in understanding the
soft Pomeron structure within the framework of
npQCD\cite{LN,KL,KKL,SHZ,JANIK,NACHT,SU,KLT,FIIM}, we are still
unable to predict the parameters associated with the
phenomenological Pomeron\cite{DL}, and to create a theory for soft
scattering. Consequently, no theoretical approach exists which
allows us to calculate a SP value to a specific given accuracy. As a
substitute, the models which were developed utilize relatively
simple soft Pomeron parametrizations and the rescattering process is
approximated by eikonal-type  models. These models satisfy  the
general principles of unitarity, including the Froissart and Pumplin
bounds, and allows us to analyze the experimental data in accordance
with these general principles.

At present, we can obtain only phenomenological estimates for SP.
Although our calculations are applicable to any small size colorless
system e.g. the production of Higgs mesons, in this paper we refer
only to the particular case of a di-jet system, as there is
experimental data from the Tevatron and HERA. In the simplest
approach we have three steps in calculating the central hard
exclusive production of a di-jet, $\,N+N \rightarrow
N+LRG+JJ+LRG+N$, which is illustrated in \fig{lrg}.

\FIGURE[ht]{
\centerline{\epsfig{file=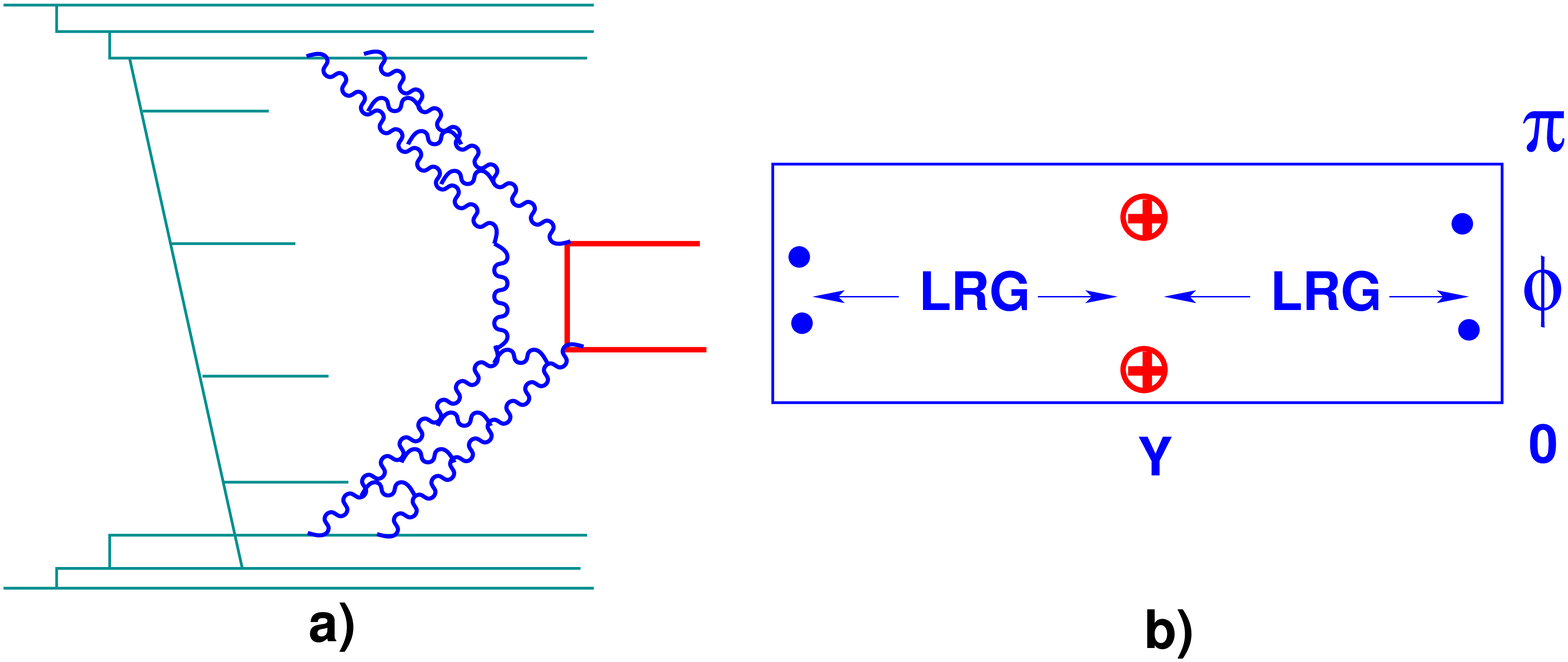,width=120mm}}
\caption{\it The contaminating soft rescatterings between partons,
that produce secondaries which may fill the rapidity gaps
(\fig{lrg}-a), and the lego-plot (\fig{lrg}-b) of
our process. The probability that there
are no additional soft interactions shown in \fig{lrg}-a, gives
the survival probability for the diffractive di-jet production.}
\label{lrg}
}

1) A description of the rescattering soft interactions.
In the eikonal model,
the elastic high energy amplitude is pure imaginary at
high energy, and can be written in impact parameter
b-space in the form
(see, for example, Refs.\cite{GLM,GLMSP})
\be
a_{el}\Lb s; b \Rb \,\,&
=&\,\,i\,\Lb \,1\,\,-\,\,e^{ - \h \Omega(s,b)}\,\Rb,\,\,\label{EM1}\\
G_{in}\Lb s; b \Rb\,\, &=&\,\,\,1\,\,-\,\,e^{-\Omega(s,b)}.\,\,\label{EM2}
\ee
$s = W^2$, where $W$ is the c.m. energy of the initial hadronic collision,
and $b$ is its impact parameter.
For the opacity $\Omega$, we use the exchange of a soft Pomeron with a
$t=0$ linear trajectory intercept of $(1+\Delta)$ and a slope
$\alpha^{,}_P$.
Accordingly,
\beq \label{OPACITY}
\Omega(s,b)\,\,=\,\,\sigma_0\,\Lb \,\frac{s}{s_0}\,\Rb^{\Delta}\,S\Lb s,b\Rb.
\eeq
$S(s,b)$ denotes the b-space normalized soft profile function satisfying
\beq \label{NORM}
\int\,\,d^2 b\,S(s,b) \,=\,1.
\eeq
In the framework of the eikonal model $\sigma_0$, $s_0$ and the
form of the profile function are phenomenological inputs,
extracted from the experimental data.
\newline
2) We need to find the impact parameter behaviour of the
two hard Pomerons, that
produce the di-jet in the central rapidity region, as is shown in
\fig{lrg}-a. We denote this
observable by $\frac{d^2 \sigma_H}{d^2 b}$. We can use the HERA DIS
data to find this dependence. In this paper we
will use the approach developed in Ref.\cite{KOTE},
in which the b-dependence of the hard Pomeron
has been extracted from the diffractive production
of $J/\Psi$ meson in the HERA DIS experiments.
\newline
3) The third step is an actual calculation of the survival
probability\cite{BJ,GLMSP}
\beq \label{SP}
<|S^2|>\,\,=\,\,\frac{\int\,\,d^2\,b\,\,\Lb d \sigma_H/d^2 b \Rb \,\,\,\,
e^{-\Omega(s,b)}\,}{\int\,\,d^2\,b\,\, \Lb d \sigma_H/d^2 b \Rb}.
\eeq
The interpretation of \eq{SP} derives from \eq{EM2}, where the factor
$e^{-\Omega(s,b)}$ secures that no
inelastic interactions occur which could change the LRG lego-plot
of \fig{lrg}-b. However, the simple
form of \eq{SP} is valid only for a single channel eikonal model
in which only elastic rescatterings are considered. It needs to
be modified in a more elaborate approach,
such as two or three channel models, in which both elastic and
diffractive rescatterings are included\cite{GLMLRG}.

The above appealing sketch of the eikonal approach demonstrates, as well,
the deficiencies which are inherent in
all presently available models, even those which are more complicated
than the eikonal rescattering approximation.
As it stands, the first and the third steps are stricly
phenomenological, and only the second step is based on well
established pQCD. This shaky theoretical situation has been the
main reason why we were not active in this field over the past five
years. The strong current demand to
evaluate the LHC diffractive cross sections and their SP, encouraged
us to return to this subject.

This paper has two main goals. To begin with, we wish to calculate a range
of possible values for exclusive LHC central di-jet (or Higgs) production.
To this end, we employ the eikonal model, being simple and transparent, so
that all uncertainties can be recognized and discussed.
We wish to avoid more complicated calculations in which
it is difficult to separate
uncertainties in the values of the input parameters, from the qualitative
behaviour of the amplitude (mostly as a function of the
impact parameter b).
We shall present and compare results stemmimg from a single channel
eikonal model,
utilizing two profile functions.
A constituent quark model (CQM) recently suggested\cite{BOLE},
and a two channel eikonal model\cite{GLMLRG}.

Our second main goal is to assess the t-dependence of the
calculated SP. The experimental study of this
differential observable requires, obviously, a significantly
higher statistics than presently available.
However, as we shall see, the discovery potential
associated with this information justifies
both experimental and theoretical efforts.

In our approach we use three principles which alleviate
our poor knowledge on soft scattering theory.
\newline
1) We aim to successfully describe the soft scattering data associated
with our investigation. This includes the data base with which we adjust
the input information for the construction of the eikonal opacity
(or opacities)
and the model predictions that can be tested experimentally.
\newline
2) Utilize all possible theoretical approaches to control our
phenomenology. In particular, the pQCD calculation for the
production of hard di-jets at short distances.
\newline
3) Even though the LHC is not yet running, we should
present our overall LHC predictions so as to be a  subject
for an experimental test even at the preliminary runs,
well before the survival probabilities
can be experimentally assessed.

For a recent review on SP, see Ref.\cite{heralhc}.

\section{Formulation}
In the following we present the main formulae pertinent to our
investigations.
\subsection{Basic formulae}
Our elastic t-channel scattering amplitude is normalized so that
\beq \label{ET}
\frac{d \sigma}{dt} =\pi\; \mid f_{el}(s,t) \mid^{2},
\eeq
\beq \label{ETT}
\sigma_{tot} = 4 \pi\,Im\,f_{el}(s,0).
\eeq
The amplitude in impact parameter space is given by
\beq \label{EI}
a_{el}(s,b) = \frac{1}{2 \pi} \int d^{2}q e^{-i \vec{q} \cdot \vec{b} }
f_{el}(s,t),
\eeq
where, $t\,=\,-q^2$.
In this representation we have
\be
\sigma_{tot}\,\,\,&=&\,\,\,2\,\int\,d^{2}b\,\,\,\mbox{Im}\,
a_{el}(s,b)\,,\label{EI1}\\
\sigma_{el}\,\,\,&=&\,\,\, \int d^{2}b\,\,\mid a_{el}(s,b)\mid^{2}.\label{EI2}
\ee
s-channel unitarity implies that $\mid a(s,b)\mid \leq 1$. When this is
written in a diagonalized form we have
\beq \label{UNT}
2\,\,\mbox{Im}\,a_{el}(s,b)\,\,=\,\,\mid a_{el}(s,b)\mid^{2}\,+\,G_{in}(s,b).
\eeq
The corresponding inelastic cross section is written
\beq \label{EI4}
\sigma_{in}\,\,=\,\,\int\,d^{2} b\,\,\,G_{in}(s,b).
\eeq
$s$- channel unitarity is most easily enforced in the eikonal approach, where
\eq{EM1}, \eq{EM2} and \eq{OPACITY} satisfy the unitarity
constraint of \eq{UNT}.

\subsection{The general expression for the survival probability}
The amplitude for central hard di-jet production is shown in
\fig{lrgdiag}.
$A_H$ denotes the exchange of a hard Pomeron
responsible for the production of the two gluonic jets (or Higgs),
at short distances.
The amplitude $A_S$ includes all possible initial state interactions due
to the exchange and interaction of soft Pomerons, including the
possibility that the two initial nucleons do not interact.
\FIGURE[ht]{
\centerline{\epsfig{file=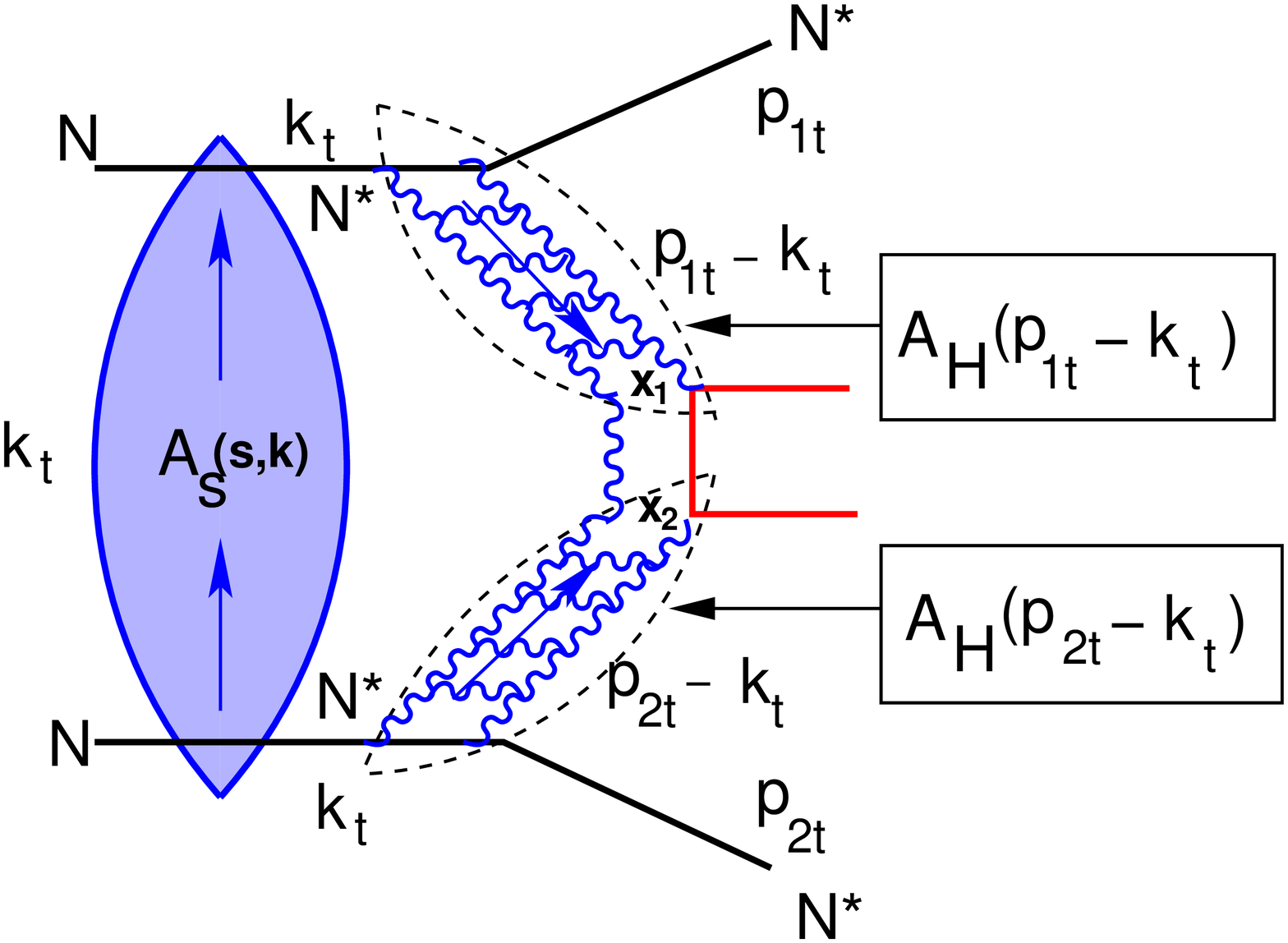,width=120mm}}
\caption{\it Central production of two hard jets
separated by two large rapidity gaps from the accompanied final state
nucleons and/or diffracted excited states (both are denoted by
$N^*$). Possible rescattering of the initial two nucleons ($N$) is
included.}
\label{lrgdiag}
}

The expression for the exclusive di-jet production (\fig{lrgdiag})
has the form
\beq \label{F1}
A \Lb N + N \,\,\to\,\,N\,+\,LRG\,+\,JJ\,+\,LRG\,+\,N \Rb\,\,=
\eeq
$$
\,\,a_{hard}\,\int\,d^2\,k_t\,
T_H\Lb (\vec{p}_{1t} - \vec{k}_{t})^2 ;x_1 \Rb\,\,T_H\Lb (\vec{p}_{2t}
- \vec{k}_{t})^2 ;x_2 \Rb
\,\,A_S\Lb k^2_t \Rb\,\,=\,\,
$$
$$
=\,\,A_H\,\,\int\,d^2\,b_1\,d^2b_2\,
e^{i\,\vec{p}_{1t}\,\cdot\,\vec{b}_1\,+\,i\,\vec{p}_{2t}\,\cdot\,\vec{b}_2}
A_H(b_1)\,A_H(b_2)\,\,A_S \Lb (\vec{b}_1\,+\,\vec{b}_2 )^2 \Rb.
$$
$a_{hard}$ is the amplitude for two hard Pomerons
fusion into two jets. This
process occurs at short distances, and we assume it to be independent
of the impact parameters.
$A_H$ is the amplitude for the hard Pomeron exchange,
which is well known in pQCD\cite{BFKL,DGLAP}.

The full hard amplitude has a more complicated dependence and can be written
as
\beq \label{TH}
T_H(p^2_t,x)\,\,=\,\,A_H(p^2_t)\,T_H(x)\,\,\longrightarrow\,A_H(b)\,T_H(x),
\eeq
where the amplitude $T_H(x)$ absorbs the energy and transverse momenta
dependence of the produced di-jet. We do not need to know the explicit
dependence of these amplitude on energy and
transverse momenta for our calculation of SP.

For the completeness of our presentation we introduce the factor
$A_{hard}$
which has been calculated in Ref.\cite{DUR2J}.
This factor is given by
\beq \label{AHARD}
A_{hard}(HP + HP \rightarrow JJ)\,\,=
\,\,\frac{32\pi}{9}\int^{P^2_T} \as(Q^2)\,\frac{dQ^2}{Q^4}
\,\phi(x_1,Q^2)\,\phi(x_2\,Q^2)\,\,e^{ - S(Q^2,P^2_T)}.
\eeq
$HP$ denotes the hard Pomeron, $\phi$ is the
unintegrated structure function and
$\vec{P}_T\,=\,\frac{1}{2}(p_{1,t} + p_{2,t})$, where $p_{i,t}$ are
the transverse momenta of the produced gluons.
The Sudakov factor $e^{-S}$ secures that there
is no radiation of gluons, with sufficiently large transverse
momenta, from the exchanged gluons in \fig{lrgdiag}.
This suppression might be cosidered as a contribution
to SP stemming from the hard sector. However,
we note that it is included in the pQCD calculation of the hard process.
See Ref.\cite{DUR2J} for detailed calculations
of $S$,$\phi$ and for the kinematical
constraints for the process of di-jet production.

The most important characteristic of the hard Pomeron exchange
for the calculation
of the SP, is its dependence on the impact parameters.
This dependence
stems from the $b$-dependence of the proton-hard Pomeron vertex which can
be defined in either BFKL\cite{BFKL} or DGLAP\cite{DGLAP} approach.
As we have mentioned, this dependence was extracted, in our calculation,
from the DIS data\cite{KOTE}, and we shall discuss it below.

$A_S$ is the amplitude accounting for the initial soft interaction,
about which little is known.
As stated, we start with a calculation of this amplitude in a
single channel eikonal model.
Accordingly, we consider only the elastic rescattering of the two initial
nucleons, neglecting the summation over the
possible excited nucleon states in \fig{lrgdiag}.

For this cross section we have
\beq
\label{F2} \frac{d \sigma}{d^2 p_{1t}\,d^2\, p_{2t}, d y_1\,d y_2}
\Lb N + N \,\,\to\,\,N\,+\,LRG\,+JJ\,+\,LRG\, + N \Rb\,\,=\,\,
\eeq
$$
=\,\,\sigma_{hard}\,\,|\int\,d^2\,b_1\,d^2b_2\,
e^{i\,\vec{p}_{1t}\,\cdot\,\vec{b}_1\,+\,i\,\vec{p}_{2t}\,\cdot\,\vec{b}_2}
A_H(b_1)\,A_H(b_2)\,\,A_S \Lb (\vec{b}_1\,+\,\vec{b}_2 )^2 \Rb|^2,
$$
where $p_{it}$ and $y_i$ are the transverse momenta and rapidities
of the produced gluons (jets).

Integrating \eq{F2} over $p_{1t}$ and $p_{2t}$, we obtain the formula for SP
(\eq{SP}) in the form
\beq \label{SPG}
\SP\,\,\,=\,\,\,\frac{\int\,d^2\,b_1\,d^2\,b_2\,\,
\left[A_H(b_1)\,A_H(b_2)\,\,A_S \Lb
(\vec{b}_1\,+\,\vec{b}_2 )^2
\Rb \right]^2}{\int\,d^2\,b_1\,d^2\,b_2\,\,
\left[A_H(b_1)\,A_H(b_2)\right]^2}.
\eeq
Let us introduce
\beq \label{F3}
\frac{d \sigma_H}{ d^2\,b} \,\,\equiv\,\,
\int\,d^2\,b'\,\,A^2_H\Lb \vec{b} - \vec{b}^{\,'}\Rb\,A^2_H(b'),
\eeq
and
\beq \label{F4}
e^{- \,\Omega(s,b)}\,\,\,\equiv\,\,A^2_S (b).
\eeq
We find that \eq{SPG} coincides with \eq{SP}. We shall show below
that \eq{F4} is a consequence of the eikonal model.

In the eikonal approximation the rescattering of two nucleons can
be described by the set of diagrams presented in \fig{aseik}.
\FIGURE[ht]{
\centerline{\epsfig{file=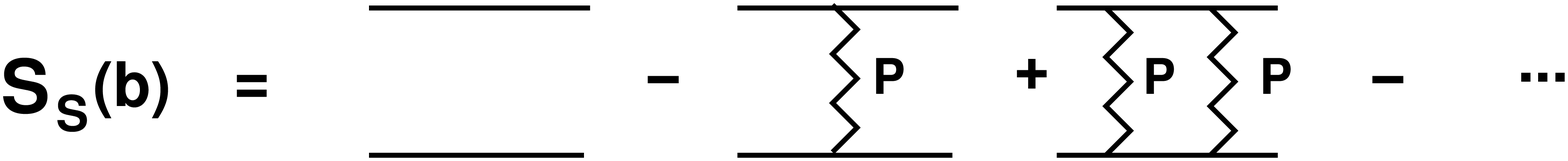,width=160mm}}
\caption{\it $A_S(b)$ in the eikonal model as a multi Pomeron exchange
chain. ${\bf P}$ denotes the exchange of a soft Pomeron.}
\label{aseik}
}
To understand the minus sign in front of the first rescattering term,
one needs to substitute this diagram into \eq{F2}, and remember
that the single Pomeron exchange, as well as the
diffractive production of jets,
are pure imaginary. This leads to a negative contribution.
See, for example, Ref.\cite{SOFT}. Summing all diagrams in
\fig{aseik}, we obtain the eikonal formula for
\beq \label{EF}
A_s(b) \,\,\,=\,\,e^{ - \h \,\Omega(s,b)},
\eeq
which leads to \eq{F4}.
It is important to realize, in this context, that different LRG di-jet
configurations, i.e. JGJ, GJJ and GJJG (which is discussed in this paper),
have different $A_s(b)$ and intermediate states
(see \fig{lrg2c} and \eq{2C9} below).
Consequently, they have different SP values.

\begin{boldmath}
\subsection{ $A_H(b)$}
\end{boldmath}
To fix the impact parameter dependence of the hard amplitude, we
return to the model of Ref.\cite{KOTE}.
The t-dependence of the cross section for the diffractive
production of $J/\Psi$ is shown in the diagram of \fig{psidd}.
\FIGURE[ht]{
\centerline{\epsfig{file=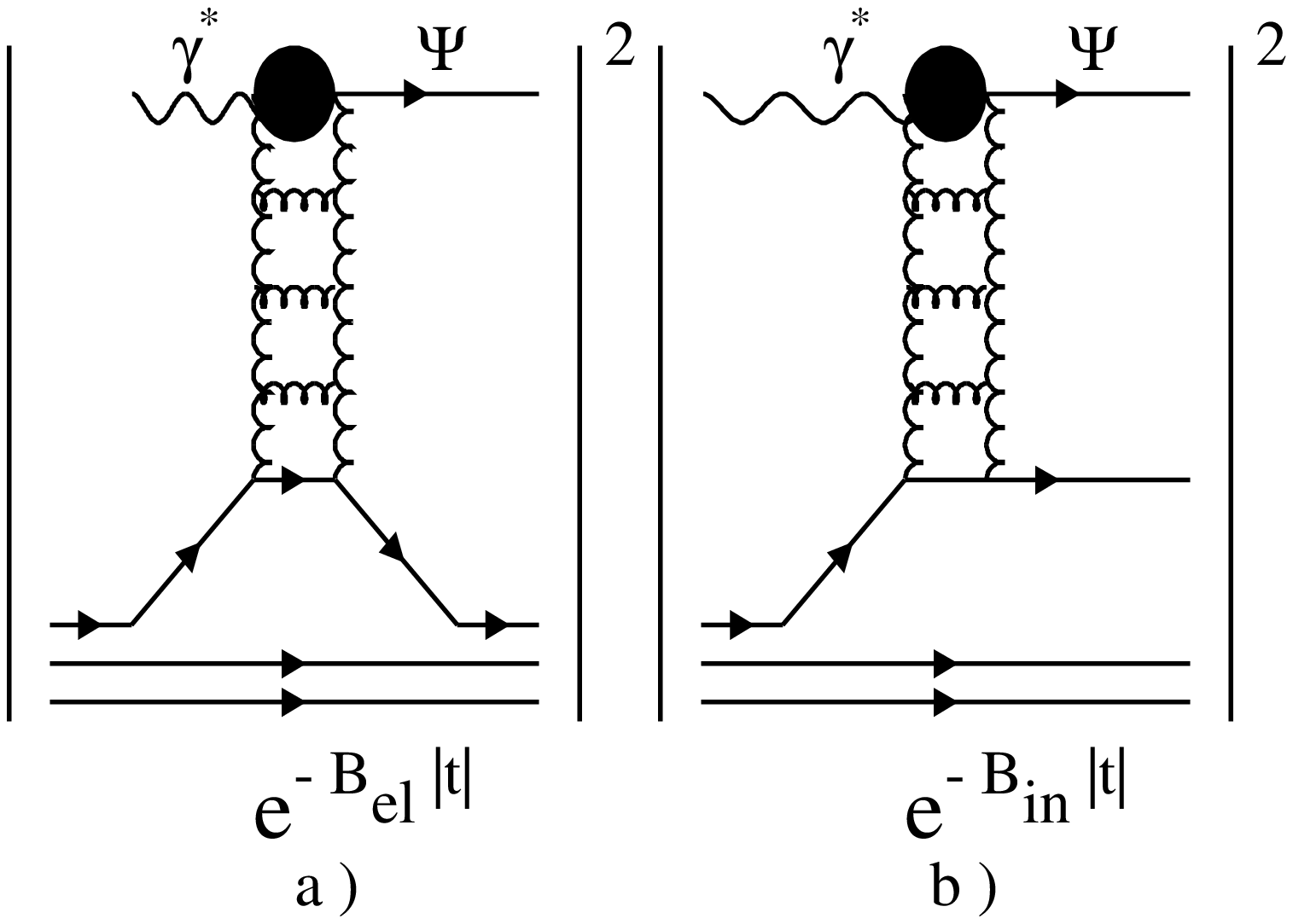,width=90mm}}
\caption{\it t-dependence of $J/\Psi$ production in DIS.}
\label{psidd}
}
The vertex for interaction of the hard Pomeron with the virtual
photon leading to the production of $J/\Psi$, can be calculated in
pQCD. From this calculation we can find the impact parameter
dependence of the proton-hard Pomeron vertex. This
vertex can be parameterized as an exponential in t
\beq \label{VNP}
A_H(t)\,\,\,=\,\,e^{ - \h B_{GG}\,|t|},
\eeq
with $B_{GG} \,\,=\,\,3.6\,\gevms$\cite{KOTE}.
The corresponding b-space transform of \eq{VNP} is
\beq\label{VNPB}
A_H(b) \,\,\,=\,\,\,\frac{1}{\pi\,R^2_H}\,\exp \Lb -\frac{b^2}{R^2_H} \Rb,
\eeq
with $R^2_H\,=\,7.2\,\gevms$. In all
our calculations below we take this form of $A_H(b)$.

Substituting \eq{VNPB} in \eq{F3} we obtain
\beq \label{DSB}
\frac{d \sigma_H}{ d^2\,b}\,\,=
\,\,\frac{1}{\pi \,R^2_H}\,\,\exp \Lb -\,\frac{b^2}{R^2_H} \Rb,
\eeq
and
\beq \label{DSB1}
\int\,d^2\,b'\,\,A_H\Lb \vec{b} - \vec{b}^{\,'}\Rb\,A_H(b')\,\,\,=
\,\,\frac{2}{\pi \,R^2_H}\,\exp \Lb -\,2\,\frac{b^2}{R^2_H} \Rb.
\eeq

The uncertainties in the determination of $R_H$ are  mostly
related to the influence of the
excited nucleons in the final state (see \fig{lrgdiag}). To
understand better the possible influence of these states on the value of SP,
we consider the process of diffractive dissociation of the
virtual photon to a $J/\Psi$ in CQM.
In this model the hard Pomeron interacts
with the constituent quark as it is shown in \fig{lrgdiag}-b.
The vertex of the hard Pomeron for this
interaction is
\beq \label{VAQM}
V_{p \to 3 q}\,\,=
\,\,\int\,\prod_i d^2 r_i \Psi_N(r_1,r_2,r_3)\,e^{- q\,x_1} \,\Psi_q(r_i),
\eeq
where $\Psi(r_i)$ is the plane wave for the quark,
and $\Psi_N$ is the wave function of the nucleon.
This wave function actually depends on two variables.
For example $r_{1,2} = r_1 - r_2$ and
$r_{3,1,2}=r_3 - \h(r_1 + r_2)$. The elastic differential cross section is
proportional to
\beq \label{AQM1}
|V_{p \to 3 q}(t)|^2\,\,=\,\,\int\,d^2 r_{1,2}\,d^2
r_{3,1,2}\,|\Psi_N(r_{1,2},r_{3,1,2})|^2\,e^{i
\vec{q}_t \cdot \vec{r}_{1,2}}
\,\,\to\,\,
\eeq
$$
1\,\,+\,\,\,q^2\,\int\,d^2 r_{1,2}\,d^2
r_{3,1,2}\,r^2_{1,2}|\Psi_N(r_{1,2},r_{3,1,2})|^2\,\,+
\,\,O(q^4)=\,\,1\,+\,q^2\,< | r^2_{1,2} | >\,\,+\,\,O(q^4).
$$
The value of $< | r^2_{1,2} | > $ can be evaluated
from the $q = \sqrt{-t}$ behaviour of the elastic diffraction
(see \fig{lrgdiag}).
This diagram leads to $A_H(t)$ of \eq{VNP} given in
the form
\beq \label{VAQM2}
A_H(t)\,\,=\,\,\int\,d^2 r_{1,2}\,d^2
r_{3,1,2}\,|\Psi_N(r_{1,2},r_{3,1,2})|^2\,e^{i\frac{2}{3}
\vec{q}_t \cdot \vec{r}_{3,1,2}}\,\,\to\,\,1\,\,+\,\,\frac{4}{9}\,
< | r^2_{1,2} | >\,\,+\,\,O(q^4).
\eeq
Therefore, $B_{GG}\,\,=\,\,\frac{8}{9}< | r^2_{1,2} | >$ .
Adding the combinatorial factors,  we see
that the
t-dependence of
the differential cross section for diffractive $J/\Psi$ production is  given by
\beq \label{SPSIAQM}
\frac{d \sigma( \gamma^* + p \to J/\Psi + 3 q)}
{d t}\,\,\propto\,\,6\,\exp\Lb - \frac{9}{8}\,B_{GG}\,|t|\Rb
\,\,\,+\,\,3\,\exp\Lb - \,B^{in}_{GG}\,|t|\Rb,
\eeq
where the second term originates from the elastic rescattering off one
quark. Using the same procedure to extract the
slope $B^{in}_{GG}$ from $B_{in}$ in \fig{psidd},
we estimate $B^{in}_{GG}\,\,=\,\,1\,\,\gevms$.

\eq{DSB} has, thus, the form
$$
\frac{d \sigma_H}{ d^2\,b}\,\,=
\,\,\frac{1}{9}\,\Lb \,4\,\frac{1}{\pi \,2\,B_{GG}}\,\,\exp \Lb
-\,\frac{b^2}{2B_{GG}} \Rb\,\,+
\,\,4\,\frac{1}{\pi \,\,(B_{GG}\,+
\,B^{in}_{GG})}\,\,\exp \Lb-\,\frac{b^2}{B_{GG}\,\,+\,\,B^{in}_{GG}}\Rb\,\,+
\right.
$$
\beq \label{DSB2}
\left.
\,\,\frac{1}{\pi \,2\,B^{in}_{GG}}\,\,
\exp\Lb-\,\frac{b^2}{2B^{in}_{GG}}\Rb \Rb.
\eeq

\subsection{Modelling the impact parameter dependence of
the soft scattering amplitude}
In the eikonal model the opacity $\Omega$ corresponds to a single soft
Pomeron exchange.
The  t-dependence of this contribution is
\beq \label{V1}
\Omega\,\,=\,\,\sigma_0 \,V^2_N(t)\,
\Lb \frac{s}{s_0} \Rb^{\Delta\,-\,\alpha'_P\,|t|},
\eeq
where $V_N(t)$ is the proton-soft Pomeron vertex with the
normalization $V(t =0) =1$.

We will use two simple models for the t-dependence of
the vertex $V_N(t)$.
\newline
1) An exponential parametrization valid in the forward t-cone.
\beq \label{EP}
V_N(t)\,\,=\,\,\exp \Lb - \frac{B_{0,el}}{4} \,|t| \Rb,
\eeq
where $B_{0,el}$ is the exponential slope of the
elastic differential cross section
due to a soft Pomeron exchange at $s = s_0$.
Although there is no theoretical justification for it, the
parametrization suggested in \eq{EP}, is in accord with all
experimental observables, that are sensitive to the small t region.
In the following it will be denoted GP.
\newline
2) A parametrization more suitable to cover a wider t range,
including the diffractive dips observed in $\frac{d\sigma_{el}}{dt}$ is
\beq \label{AQM}
V_N(t)\,\,=\,\,G_N(t).
\eeq
$G_N(t)$ is the proton
electromagnetic form factor (see for example Ref.\cite{DL}).
We will use the dipole parameterization for this form factor
\beq \label{G}
G_N(t)\,\,\,=\,\,\,\frac{1}{(1 + t/m^2)^2}\, ,
\eeq
with $m^2 = 0.72\,\gevs$.
The above parametrization is used in CQM.

The profile functions corresponding to GP and PP have the form
\be \label{S1}
S_{GP}(s,b)\,\,&=&\,\,\frac{1}{2 \pi\,B_{el}}\,\exp \Lb
- \frac{b^2}{2 B_{el}}\Rb,
\ee
with $B_{el} \,\,=\,\,B_{0,el}\,\,+\,\,2\,\alpha'_P \,\ln(s/s_0)$.
\be
S_{PP}(s,b)\,\,&=&\,\,\,\int\,q\,d \,q\,J_0 \Lb q\,b \Rb
\,G^2_N(t)\,\,\,\exp \Lb -\,\,\alpha'_P\,\ln(s/s_0) \,\,\,q^2\,\Rb.
\label{S2}
\ee

\subsection{Explicit analytic formulae for the Gaussian parametrization}
Using the parameterization of \eq{EM1}, \eq{EM2}, \eq{OPACITY} and
\eq{S1}) for the soft profile function,
we can obtain simple analytic expressions for the
main observables of the soft interaction
\cite{GLMSP}.
\be
\sigma_{tot} & = & 2 \int d^{2}b (1 - e^{- \Omega(s,b)/2})
= 4 \pi B_{el}(s)\left[ln(\frac{\nu(s)}{2}) + C - Ei(- \frac{\nu(s)}{2})
\right],
\label{AF1}\\
\sigma_{in} &=&  \int d^{2}b (1 - e^{- \Omega(s,b)})
= 2 \pi B_{el}(s) \left[ln(\nu(s)) + C - Ei(- \nu(s))\right],
\label{AF2}\\
\sigma_{el} &=& \sigma_{tot} - \sigma_{in}
=2\pi B_{el}(s) \left[ln(\frac{\nu(s)}{4})+C + Ei(-\nu(s))-
2 Ei(-\frac{\nu(s)}{2}) \right],
\label{AF3}\\
\sigma_{el}/\sigma_{tot} &=&
\h \left[ln(\frac{\nu(s)}{4})+C+Ei(-\nu(s))-2Ei(-\frac{\nu(s)}{2})\right]/
\left[ln(\frac{\nu(s)}{2})+C-Ei(-\frac{\nu(s)}{2})\right],
\label{AF4}
\ee
where
\beq \label{NU}
\nu(s)\,\,=\,\,\Omega(s,b=0),
\eeq
$Ei(x)=\int_{-\infty}^{x}\frac{e^{t}}{t}dt$
and C = 0.5773. From \eq{AF4} we note that the ratio of
$\sigma_{el}/\sigma_{tot}$  depends only
on the value of $\nu(s)$ and, therefore, provides a possibility to find
the value of $\nu$ directly from the experimental data\cite{GLMSP}.

The SP, given by \eq{SP}, can also be calculated analytically if we
assume \eq{DSB} for $d \sigma_H(b)/d^2 b$. We obtain
\beq \label{AFSP}
\SP\,\,=\,\,\, \frac{a(s) \gamma \Lb a(s),
\nu(s) \Rb}{\left[\nu(s)\right]^{a(s)}}.
\eeq
The incomplete gamma function is
$ \gamma(a,x) = \int_{0}^{x} z^{a-1}e^{-z}dz $ and
$a(s)\,=\,\frac{2\,B_{el}(s)}{R^{2}_{H}(s)}$.
For the profile function of \eq{S2} we cannot provide
a set of analytic formulae. We have used the above formulae as a check
of our methods and numerical computations.

\subsection{Constituent quark model}
The characteristics of the soft processes can be alternatively calculated
also with CQM\cite{BOLE}.
In CQM we assume that all hadrons consist of
constituent quarks (two for a meson and three for a baryon), and all
scattering processes go through the interactions of the constituent
quarks. In this model the state of three free quarks diagonalizes the
interaction matrix,
the proton state, and the diffraction dissociation state,
which can be viewed as the expansion
of the three quark plane wave function
\beq \label{AQM0}
\prod_{i=1}^3\,\,\Psi_q(r_i)\,\,=
\,\,\alpha\,\Psi_p\,\,+\,\,\beta\,\Psi_D,
\eeq
with a normalization $\alpha^2\, +\, \beta^2 \,=\,1$. The
constituent quark model can be viewed, thus, as a particular realization
of a two channel model
which we will consider below on a more general basis.

For a constituent quark-quark scattering amplitude we use the
eikonal formulae \eq{EM1} and \eq{EM2}.
Accordingly, the amplitude for the scattering of a quark
$i_1$ from the first hadron with a
quark $i_2$ from the second hadron is
\beq \label{QQA}
A_{i_1,i_2}(s,b)\,\,=
\,\,i \,\left( \,1\,\,-\,\,\exp \left( - \frac{\Omega_{i_1,i_2}(s,b)}{2}
\right)\,\right).
\eeq
The opacity $\Omega_{i_1,i_2}$ is given by
\beq \label{AQM2}
\Omega_{i_1,i_2}^{CQM}\,\,=\,\,\,
\eeq
$$
\,\,\frac{\sigma_0}{\pi\,\Lb
8\,B^{in}_{GG}\,\,+\,\,4\,\alpha'_P\,\ln (s/s_0) \Rb}\,\,\,\Lb
\,\frac{s}{s_0}\,\Rb^{\Delta}\,\,\exp\Lb  -
\frac{b^2_{i_1,i_2}}{ 8\,B^{in}_{GG}\,\,+\,\,4\,\alpha'_P\,\ln (s/s_0)}\,\Rb,
$$
where $\vec{b}_{i_1,i_2}\,\,=\,\,\vec{r}_{i_1,k_1,l_1}\,\,
=\,\,\vec{r}_{i_1}\,\,-\,\,\h (\vec{r}_{k_1} + \vec{r}_{l_1})$.

In this paper we simplify \eq{AQM2} assuming that
$b\,\,\gg\,\,r_{i_1,k_1,l_1} \approx \,r_{i_2,k_2,l_2}$.
It is certainly correct at very high energies,
since $b^2\,\,\propto\,\,4\,\alpha'_P\,\ln
(s/s_0) \,\,\gg\,\,r^2_{i_1,k_1,l_1}$.
More advanced calculations in the framework of CQM
have been presented in Ref.\cite{BOLE}.

Having defined the opacities,
we obtain the following formulae for the soft cross sections
\be
\sigma_{tot}\,\,&=
&\,\,9\,\,\times\,\,2\,\int\,d^2b\,\Lb\,1\,-\,\exp\Lb\,-\h\,\Omega^{CQM}(s,b)
\Rb \Rb,
\label{AQM3}\\
d\,\sigma_{el}/d t\,\,&=
&\,\,9^2\,\,\times\,\,G^4_N(t)\,|f^{CQM}(s,t)|^2\,,
\label{AQM4}\\
d\,\sigma_{diff}/dt+d\,\sigma_{el}/d t\,\,&=
&\,\,9^2 \times\,\,\Lb \frac{2}{3}\exp\Lb -
\frac{9}{8}\,[B_{GG}\,-\,B^{in}_{GG} ]\,|t|\Rb
+\frac{1}{3}\,\,\Rb^2\,\,|f^{CQM}(s,t)|^2\,\,.
\label{AQM5}
\ee
$f^{CQM}(s,t)$ is the quark-quark scattering amplitude in the
t-representation, see \eq{ET} and
\eq{EI} for the relations between t and b
representations. $G_N(t)$
is the electro-magnetic
proton form factor (see \eq{G}).
The numerical combinatorial factors give the number of interacting
quark-quark (antiquark) pairs. Note that in the CQM diffraction
is confined to small masses\cite{BOLE}.

In CQM, with $d^2 \sigma_H/d^2 b$ given by \eq{DSB2}, we obtain
\beq
\label{SPQM}
<|S^2|>\,\,=\,\,\frac{\int\,\,d^2\,b\,\,\Lb d\sigma_H/d^2 b \Rb
\,\,\,\,e^{-9\,\cdot\Omega_{i_1,i_2}(s,b)}}
{\int\,\,d^2\,b\,\,\Lb d\sigma_H/d^2b\Rb},
\eeq
where $\Omega_{i_1,i_2}$ is determined by \eq{AQM2}.
\eq{SPQM} determines the probability
that none of the possible nine
quark-quark pairs can rescatter inelastically
leading to the production of additional hadrons.

Comparing the calculations using this model with the
calculations using the eikonal approach, we
are able to assess how
sensitive the value of SP is to a more elaborate model.

\subsection{Two channel model}
A major deficiency of the single channel eikonal model is its failure
to reproduce
the observed mild energy dependence of $\sigma_{sd}$ in the ISR
and above\cite{GLM}. In a single channel model, $\sigma_{sd}$ is assumed
to be small enough so that
$\frac{\sigma_{sd}}{\sigma_{el}}$ is a small parameter. Hence, the eikonal
rescatterings are elastic. Accordingly, $\sigma_{sd}$ is not included in
the fitted data base fixing the single channel model parameters.
Regardless, the SP of $\sigma_{sd}$ can be calculated resulting with
disappointing output.
The model's inability to reproduce the diffractive cross section energy
dependence reflects an over estimation of SP at higher energies
(see Ref.\cite{GLM} for details).

To overcome this difficulty we have developed a more
elaborate multi channel
eikonal model\cite{GLMLRG} in which both elastic and diffractive soft
rescatterings are included resulting in a decrease of
the calculated high energy SP.
In the two channel version of the model double diffraction
is assumed to be small enough so as to be neglected in the eikonal
rescatterings. This assumption is supported by the data available in the
ISR-Tevatron range.
Note that
the addition of a competing diffractive channel actually reduces the
elastic screening. In our model, the added
diffractive screening
adds more screening than the loss it has induced.
The final results in an over all larger screening which
reproduces the SD data well, considering the scatter of the experimental
points\cite{heralhc}.

In a two channel model, the two hadron states of a proton and
a diffractive state are considered simultaneously.
\be
\Psi_{p}\,&=& \,\alpha\,\Psi_1\,\,+\,\,\beta\,\Psi_2\,\,, \label{2C1} \\
\Psi_{D}\,&=& -\,\beta\,\Psi_1\,\,+\,\,\alpha\,\Psi_2
\,\,.
\label{2C2}
\ee
p and D denote the proton (\eq{2C1}) and diffractive (\eq{2C2})
states. The wave functions functions $\Psi_1$ and $\Psi_2$ are
orthogonal. They diagonalize the interaction matrix at high
energies and, therefore,
\beq \label{2C3}
A^{i \rightarrow j}_{k\rightarrow l} \,\,=
\,\,A_{i,k}\,\delta_{i,j}\,\delta_{k,l}\,.
\eeq
As in the single channel formulation
\beq \label{2C4}
A_{i,k}\,\,=\,\, i \left(\,1\,\,-\,\,e^{ - \h \Omega_{i,k}} \,\right).
\eeq
For the opacities
$\Omega_{i,k}(b)$ we use a GP parameterization
\beq \label{2C5}
\Omega_{i,k}(b)\,\,=\,\,\nu_{i,k}\,e^{-\frac{b^2}{R^2_{i,k}}}\,=
\,\frac{\sigma^0_{i,k}}{\pi\,R^2_{i,k}(s)}
\,\left(\frac{s}{s_0}\right)^\Delta\,\,e^{-\frac{b^2}{R^2_{i,k}}},
\eeq
where
$\left(\sigma^0_{i,k}\right)^2\,=\,\sigma^0_{i,i}\,\sigma^0_{k,k}$
and
\beq\label{RIK}
R^2_{i,k}(s)\,=\,2\,R^2_{i,0}\,+\,2 \,R^2_{k,0}\,\,+
\,\,4\,\alpha'_P\,\ln(s/s_0).
\eeq
More details regarding the parametrization can be found in
Ref.\cite{GLMLRG}.
An important fitted parameter which is essential for the present study is
$\beta = 0.464$.

For the calculation of exclusive di-jet central production SP
we need to generalize \eq{SP} for a two channel scenario.
This generalization is illustrated in \fig{lrg2c}, leading to a
generalization of \eq{F1}
\FIGURE[ht]{
\centerline{\epsfig{file=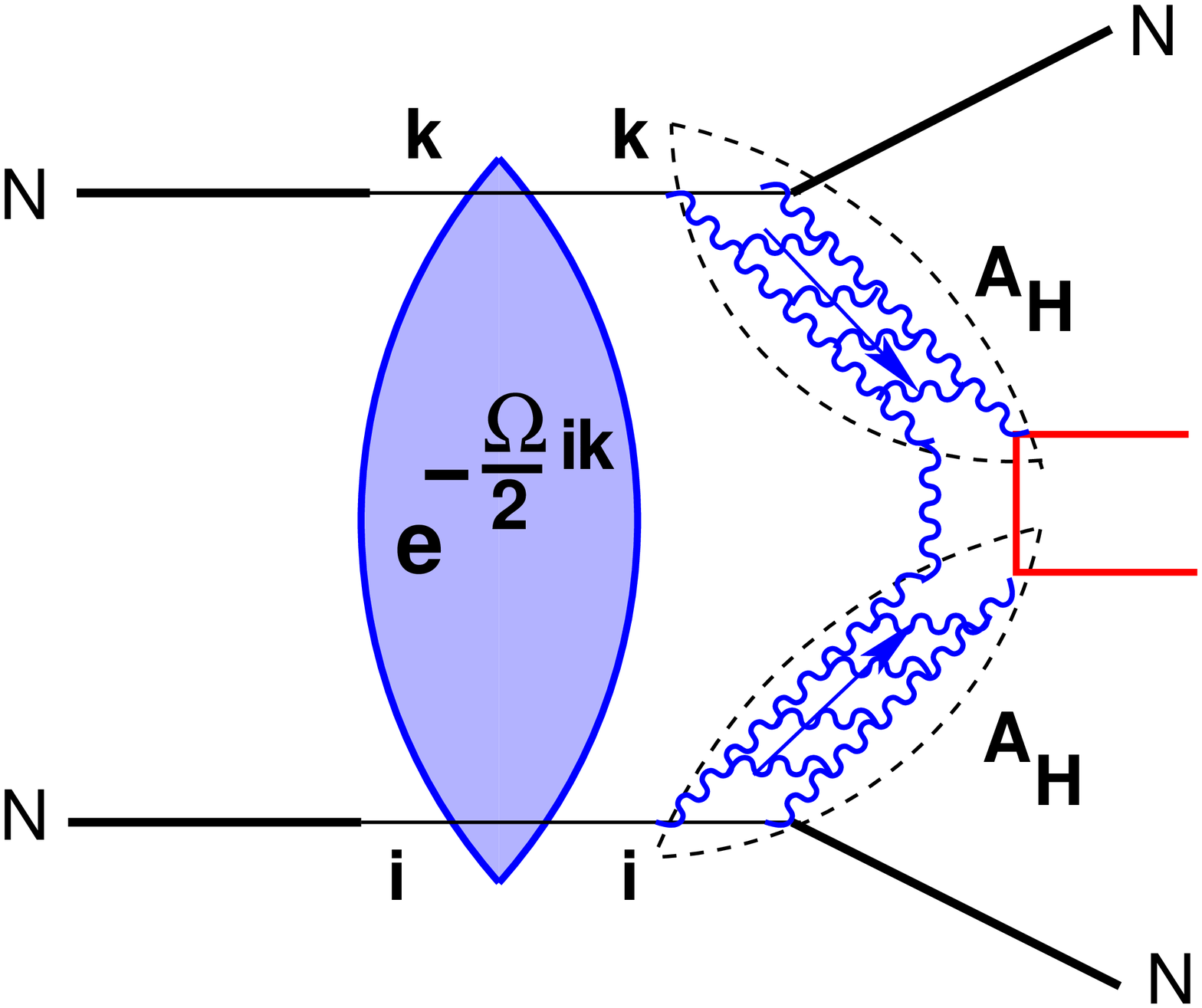,width=90mm}}
\caption{\it Production of di-jet with two large rapidity gaps in
the two channel eikonal model for the soft interaction.}
\label{lrg2c}
}
\beq \label{2C8} A\Lb N+N\,\,\to\,\,N\,+\,LRG\,+JJ+\,LRG\, + N \Rb\,\,=
\eeq
$$
=\,\,a_{hard}\,\,\sum_{i,k} \int\,d^2\,b_1\,d^2
b_2\,e^{i\,\vec{p}_{1t}\,\cdot\,\vec{b}_1\,
+\,i\,\vec{p}_{2t}\,\cdot\,\vec{b}_2}
A_H(i \to p; b_1)\,A_H(k \to p; b_2)\,\,e^{-\frac{\Omega_{i,k}
\Lb (\vec{b}_1\,+\,\vec{b}_2 )^2 \Rb}{2}}.
$$

This equation can be rewritten in an explicit way using \eq{2C1}
(see Refs.\cite{GLMLRG,heralhc} for details).
\beq \label{2C9}
A\Lb N + N \,\,\to\,\,N\,+\,LRG\,+JJ+\,LRG\, + N \Rb\,\,=
\eeq
$$
=\,\,a_{hard}\,\, \int\,d^2\,b_1\,d^2
b_2\,e^{i\,\vec{p}_{1t}\,\cdot\,\vec{b}_1\,
+\,i\,\vec{p}_{2t}\,\cdot\,\vec{b}_2}
\,\Lb \alpha^2\,e^{ - \frac{-\Omega_{1,1}\Lb(\vec{b}_1\,+
\,\vec{b}_2)^2 \Rb}{2}}\,A_H(1 \to p; b_1)\,A_H(1 \to p; b_2)\,\,\right.
$$
$$
\left. +\,\,\alpha\,\beta\,e^{-\frac{-\Omega_{1,2}
\Lb(\vec{b}_1\,+\,\vec{b}_2)^2 \Rb}{2}} \,
\{A_H(1 \to p;b_1)\,A_H(2\to p; b_2)\,+
\,A_H(2 \to p;b_1)\,A_H(1\to p; b_2)\}\right.
$$
$$
\left.\,+\, \beta^2\,\,e^{ - \frac{-\Omega_{2,2}\Lb(\vec{b}_1\,+
\,\vec{b}_2)^2 \Rb}{2}}\,A_H(2 \to p;b_1)\,A_H(2 \to p; b_2) \Rb.
$$

\eq{2C9} can be written in a more convenient form which enables  us
to use the experimental data corresponding to
\fig{psidd}. We introduce two amplitudes
\be
A_H(p \to p;b)\,\,&=&
\frac{V_{p \to p}}{2 \pi B_{GG}}\,\,\exp \Lb-\frac{b^2}{2\,B_{GG}} \Rb\,,
\label{2C10}\\
A_H(p \to D;b)\,\,&=&\,\,\frac{V_{p \to D}}{2 \pi
B^{in}_{GG}}\,\,\exp \Lb -\frac{b^2}{2\,B^{in}_{GG}}\Rb\,,
\label{2C11}
\ee
where the input parameters $B_{GG}$ and $B^{in}_{GG}$ have been
introduced in Section II.C.
The input assumption of the two channel model\cite{GLMLRG} is
that the double diffractive production is small enough to be
neglected. Using this assumption, together with \eq{2C10}, \eq{2C11},
\eq{2C1} and \eq{2C2}, enables us to rewrite \eq{2C9}
\beq \label{2C12} A \Lb N + N \,\,\to\,\,N\,+\,LRG\,+JJ+\,LRG\,+ N \Rb\,\,=
\eeq
$$
=\,\,a_{hard}\,\, \int\,d^2\,b_1\,d^2
b_2\,e^{i\,\vec{p}_{1t}\,\cdot\,\vec{b}_1\,
+\,i\,\vec{p}_{2t}\,\cdot\,\vec{b}_2}
\,\,\exp\Lb - \frac{-\Omega\Lb (\vec{b}_1\,+\,\vec{b}_2)^2
\Rb}{2}\Rb \,\,
$$
$$
\Lb
\Lb 1 -2\,\beta^2 s_D \Lb (\vec{b}_1\,
+\,\vec{b}_2)^2 \Rb \Rb\,A_H(p \to p;b_1)\,A_H(p \to p;b_2) \right.
$$
$$
\left.\,-\,\,2\,\alpha\,\beta \,\,s_D \Lb
(\vec{b}_1\,+\,\vec{b}_2)^2 \Rb \,\{ \,A_H(p \to p;b_1)\,A_H(p \to
D; b_2) \,\,+\,\,A_H(p \to D;b_1)\,A_H(p \to p; b_2) \Rb.
$$
In our notation
\beq \label{2C13}
s_D(b)\,\,=\,\,1\,\,-\,\,\exp\Lb -\frac{\Delta \Omega (b)}{2} \Rb,
 \eeq
with $\Delta \Omega (b) = \Omega_{1,2} - \Omega_{1,1}$.

For the opacities $\Omega$ and $\Delta \Omega$ we use the
forms and fitted parameter values of Ref.\cite{GLMLRG}.
\be
\Omega(b)\,\,&=\,\,\frac{\sigma_{0,p}}{\pi\,R^2_{p}(s)}\,\Lb
\frac{s}{s_0} \Rb^{\Delta}\,\exp\Lb -\frac{b^2}{R^2_p(s)}\Rb\,,
\label{2C14}\\
\Delta \Omega(b)\,\,&=\,\,\frac{\sigma_{0,D}}{\pi\,R^2_{D}(s)}\,\Lb
\frac{s}{s_0} \Rb^{\Delta}\,\exp\Lb-\frac{b^2}{R^2_D(s)}\Rb \,,
\label{@C14}\\
\ee
where
\beq \label{2CR}
R^2_{p\, or\, D}\,\,\,=
\,\,R^2_{0,p\,or\, D}\,\,\,+\,\,4\,\alpha'_P\,\ln(s/s_0) .
\eeq
Note that $R^2_{0,D}\,=\,\frac{1}{2} R^2_{0,p}$.

To obtain the SP we need to substitute the following two elements in
\eq{SP}.
\beq \label{2CSP1}
\frac{d \sigma_H}{d^2 b}\,\,\longrightarrow\,T^H_{el}(b),
\eeq
and
\beq
\label{2CSP2} e^{-\Omega}\,\frac{d \sigma_H}{d^2 b}\,\,
\longrightarrow\,\,e^{-\Omega}\,\,\Lb
(1 - 2\,\beta^2 s_D)^2 \,T^H_{el}(b)\,\,\right.
\eeq
$$
\left. -\,\,  4\,\,v\,\alpha\,\beta ( 1 -2 \,\beta^2
s_D)\,s_D\,T^H_1(b)\,\,+\,\,8\,v^2\,\alpha^2\,\beta^2\,s^2_D
\,\{\, T^H_2(b) \,+\,\,T^H_3(b) \} \Rb.
$$
$v\,=\,V_{p \to D}/V_{p \to p}$, and
\be
T^H_{el}\,\,&=&\,\frac{1}{2\,\pi\,B_{GG}}\,\exp\Lb - \frac{b^2}{2 B_{GG}}\Rb\,,
\label{2CT1}\\
T^H_{i}\,\,&=&\,\frac{1}{\pi\,B^H_{i}}\,\exp\Lb - \frac{b^2}{B^H_{i}} \Rb\,.
\label{2CT2}
\ee
The corresponding input slopes are
\be
B^H_1\,\,&=&\,\frac{1 + 3 \kappa}{1 + \kappa}\,B_{GG}\,,
\label{2CT3}\\
B^H_2\,\,&=&\,( 1 + \kappa)\,B_{GG}\,,
\label{2CT4}\\
B^H_3\,\,&=&\,\frac{ 4 \kappa}{1\,+\,\kappa}\,B_{GG}\,,
\label{2CT5}
\ee
where $\kappa = B^{in}_{GG}/B_{GG}$.

\section{Results and Comparisons}
In this section we present the  results of the four models we have
considered and compare them with the relevant experimental data. We
then proceed to present and assess our SP predictions aiming to
define a value and a
range for the SP of a central GJJG exclusive dijet production at the LHC.

\subsection {Adjusted parameters}
Following is a summary of the phenomenological adjusted
parameters for the soft input of the four models we have considered.
\be
\mbox{The intercept of the soft Pomeron trajectory at}\,\,
t = 0 &\,\,\,\,\,\, &\Delta\, ,\label{PS1}\\
\mbox{The slope of the soft Pomeron trajectory at} \,\,
t = 0 & & \alpha'_P\, ,\label{PS2} \\
\mbox{The initial energy squared} & & s_0 \,, \label{PS3} \\
\mbox{The strength of the Pomeron interaction at} \;
s = s_0\,\, & &\sigma_0 \,, \label{PS4}\\
\mbox{The slope of the vertex} \; V_N(t)\,\, & & B_{0,el}\,, \label{PS5}\\
\mbox{The slope of the vertices in the two channel model}
& & R^2_{0,p}\, \,\mbox{and}\,\,  R^2_{0,D}\,,\label{PS51}\\
\mbox{The fraction of}\, \Psi_D\, \mbox{and}\, \Psi_2 & & \beta\, .
\label{PS52}
\ee
We determine these parameters by fitting the value and energy
dependence of the soft data base observables. As previously mentioned, the
input
assumption of a single channel eikonal model is that
$\frac{\sigma_{diff}}{\sigma_{el}}<<1$. Accordingly, its data base
consists of $\sigma_{tot}$,
$\sigma_{el}$ and $B_{el}$, whereas
$\sigma_{sd}$ is a prediction. CQM parameters are adjusted from
the same data base. Since this model applies
only to small mass diffraction\cite{BOLE}, it cannot
predict $\sigma_{sd}$.
The data base for the two channel model includes the above and in addition
the $\sigma_{sd}$ data points.
We calculate the corresponding SP as a prediction derived after
fixing these parameters, provided we can specify the opacity (opacities)
of the screened hard process.

The best fit adjusted parameters for the models considered are
presented in Table I. Note that even though the fitted $\sigma_0$ values
obtained for the single channel model GP and PP profiles are identical,
the corresponding $\nu$ values are different reflecting different
b-distributions.
\begin{scriptsize}
\begin{sidewaystable}
\begin{center}
\begin{tabular}{| c | c | c | c | c | c |}
\hline \hline
    & & & & &              \\
Parameters   & \,\,\,\, $\Delta$\,\,\,\, &\,\,\,\, $\alpha'_P(0)$\,\,\,\,
& \,\,\,\,$s_0$\,\,\,\,
&\,\,\,\, $\sigma_0$\,\,\,\, &\,\,\,\, Slope\,\,\,\,  \\
  & & & & &              \\ \hline
Gaussian parameterization  & & & & & \\
(GP, \eq{S1}) &\,\,\,\, 0.09\,\,\,\, &\,\,\,\, 0.25$\gevms$ \,\,\,\,
&\,\,\,\, 450 $\gevs$ \,\,\,\,&\,\,\,\,
47.2\, mb\,\,\,\,  &\,\,\,\,$B_{0,el}$\,=\, 10.24\, $\gevms$\,\,\,\, \\
 & & & & & \\\hline
Power -like parameterization   & & & & & \\
(PP, \eq{S2})  &0.09 &0.25$\gevms$ &450 $\gevs$ & 47.2 \, mb
& $m^2$ = 0.72 $\gevms$\\
  & & & & &              \\ \hline
 Constituent Quark Model & & & & &              \\
 (CQM, \eq{AQM2}-\eq{AQM5}) & 0.08 & 0.28$\gevms$ & 250 $\gevs$ &  4.13 mb
&
$B^{in}_{GG}$\,=\,
0.5
$\gevms$ \\
 & & & & &              \\
\hline
Two channel model  & & & & & \\
2ChM  $\beta \,=\,0.464$  & 0.126 & 0.2 $\gevms$ & 1\, $\gevs$ &
5.07 mb ($\sigma^0_{p}$) & 16.34 $\gevms$
($R^2_{0,p}$) \\
(\eq{2C1} - \eq{2C5}) & & & & 56.5  mb ($\sigma^0_{D}$)
&  8.17 $\gevms$  ($R^2_{2,D}$)\\
 & & & & & \\
\hline \hline
\end{tabular}
\caption
{\it The best fit adjusted values of the Pomeron parameters in
the models considered.}
\end{center}
\end{sidewaystable}
\end{scriptsize}

\subsection{Reproduction of the soft scattering observables}
A quality reproduction of $\sigma_{tot}$, $\sigma_{el}$ and
$B_{el}$ is an obvious prerequisite for a soft scattering model
with which we may calculate the SP of interest.
As we noted, $\sigma_{sd}$ is not
included in the soft data base of the single channel eikonal model, but it
is a prediction of the model\cite{GLM}. CQM is not suitable to
calculate $\sigma_{sd}$. In the two channel model, $\sigma_{sd}$ is included
in the fitted data base. In the following we discuss the details
of the soft scattering output of the models we have considered.
\subsubsection{One channel model}
For the purpose of our present investigation we have considered a toy
soft DL like\cite{DL} Pomeron exchange model, neglecting the secondary
Regge exchanges. This is a strictly high energy model for which we take,
never the less, a relatively low $s_0\,=\,450\,\gevs$. This choice should
be compared with standard Regge pole parametrizations in which
the Regge contribution diminishes at much higher energies.
Consequently, this model can not be continued to lower energies.
The best adjusted Pomeron parameters are given in Table I.
Regardless of its simplicity, this model provides a very reasonable
reproduction of its data base in the ISR-Tevatron enery range.
An interesting observation is that the two
profiles result in remarkably close outputs.
This is so even though the $\nu$ values corresponding to the two profiles
are quite different. $\nu_{GP}(450) = 1.88$, while
$\nu_{PP}(450) = 2.31$. We conclude that a good reproduction of the data
base requires a delicate compensation between $\nu$ and the effective
radius of the two profiles respectively.

Checking the b-distribution output of a given eikonal model, we encounter
a fundamental feature, which is very transparent
in the single channel version with a GP profile input for both the
elastic and diffractive amplitudes.
The elastic amplitude output, which respects unitarity, maintains an
approximate Gaussian b-distribution peaking at b = 0. On the other hand,
the diffractive amplitude output is non Gaussian
peaking at a non zero b. The sum
of the two amplitudes respects the Pumplin bound. This phenomena is easily
understood once we note that the eikonalization of a diffractive
amplitude amounts to the transition
$M_{diff}(s,b) \rightarrow M_{diff}(s,b) e^{-\Omega(s,b)}$. Since
$\Omega$ is central its suppressing effect is maximal at small b, and
diminishes at high b
(For details see Ref.\cite{GLMSP,GLMLRG} and the
preprint first version of this paper
\cite{GLMTHIS}).

As noted,
the problematic asset of the single channel model is its inability to
reproduce the energy dependence of $\sigma_{sd}$, which implies an over
estimation of the corresponding SP becoming worse with increasing
energy. As a result, we consider any SP calculated within the single
channel to represent, at best, an upper limit for the SP.

\subsubsection{CQM}
The CQM does not provide a good reproduction of the soft data
base at low and medium energies. It is applicable for the Tevatron
energies and above.
Note that in this model the PP profile is naturally related
the electro-magnetic form factor of the proton.
CQM is limited to small mass diffraction only, since the
triple Pomeron vertex is not included in its formulation.
Consequently, $\sigma_{sd}$ is not included in its soft data base.
Since the model takes into account the rescatterings due to low mass
exitation of the interacting hadrons it may be classified as an example
of a simple two channel model.

The advantages of the model is it simplicity and the realistic
b-dependence.
The model does enable us a calculation of low mass
$\frac{d\sigma_{diff}}{dt}$/$\sigma_{diff}$
presented in \fig{dsddt} at LHC energy.
The positions of the calculated dips are a consequence of the input PP
profile and should be considered reasonably reliable since the model, with
this input, correctly reproduces the t dependence of the
Tevatron elastic cross sections.
\FIGURE[ht]{
\centerline{\epsfig{file=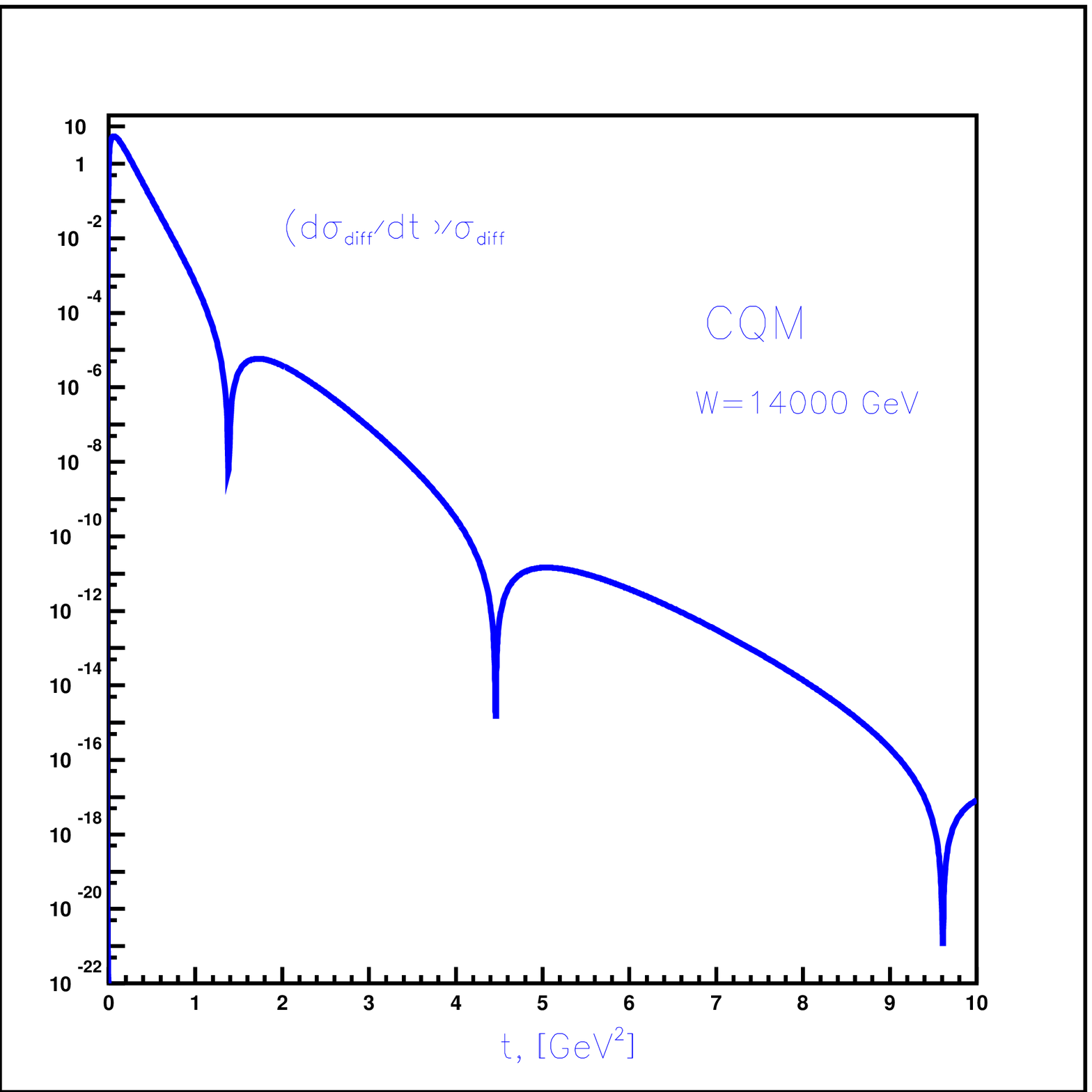,width=90mm}}
\caption{\it CQM LHC prediction for the
t-dependence of the low mass diffractive differential cross section.}
\label{dsddt}
}

\subsubsection{Two channel model}
Our reproduction of $\sigma_{tot}$, $\sigma_{el}$, $B_{el}$  and
$\sigma_{sd}$ in a two channel eikonal model have been published a while
ago\cite{GLMLRG}. Note that the soft data base includes the single
diffraction data. Our
$\frac{\chi^2}{d.o.f.}$ = 1.5, which is seemingly
high, reflects the poor quality of the $\sigma_{sd}$ points.
Note that
the b-space peripheral behaviour of the diffractive channels amplitudes
is maintained in the two channel scenario. Our LHC
predictions for $\sigma_{tot}$ = 103.8 mb, $\sigma_{el}$ = 24.5 mb,
$B_{el}$ = 20.5 $GeV^{-2}$ and $\sigma_{sd}$ = 12.0 mb.
We consider these predictions, and the consequent SP calculated values,
reliable as our input diffractive rescatterings are based on an
effective parametrization which  includes the high mass exitations.
The above numbers are significant
for central Higgs diffractive production, where there is a problematic
background of single diffraction which mimics
the sought after Higgs signal.

\subsection{Survival probabilities}
Using \eq{SP} we calculate the energy dependence of the
SP for our process in the single channel eikonal model.
The result is shown in \fig{sp}, which compares the exclusive central
di-jet SP as calculated in the four models we have considered.
For a single channel model we see that the profile
function for the soft interaction does not considerably affect
the value of the SP. The GP and PP profiles produce a small SP difference
at the ISR energies, but at higher energies, Tevatron and above, the
results are essentially the same.
\FIGURE[ht]{
\centerline{\epsfig{file=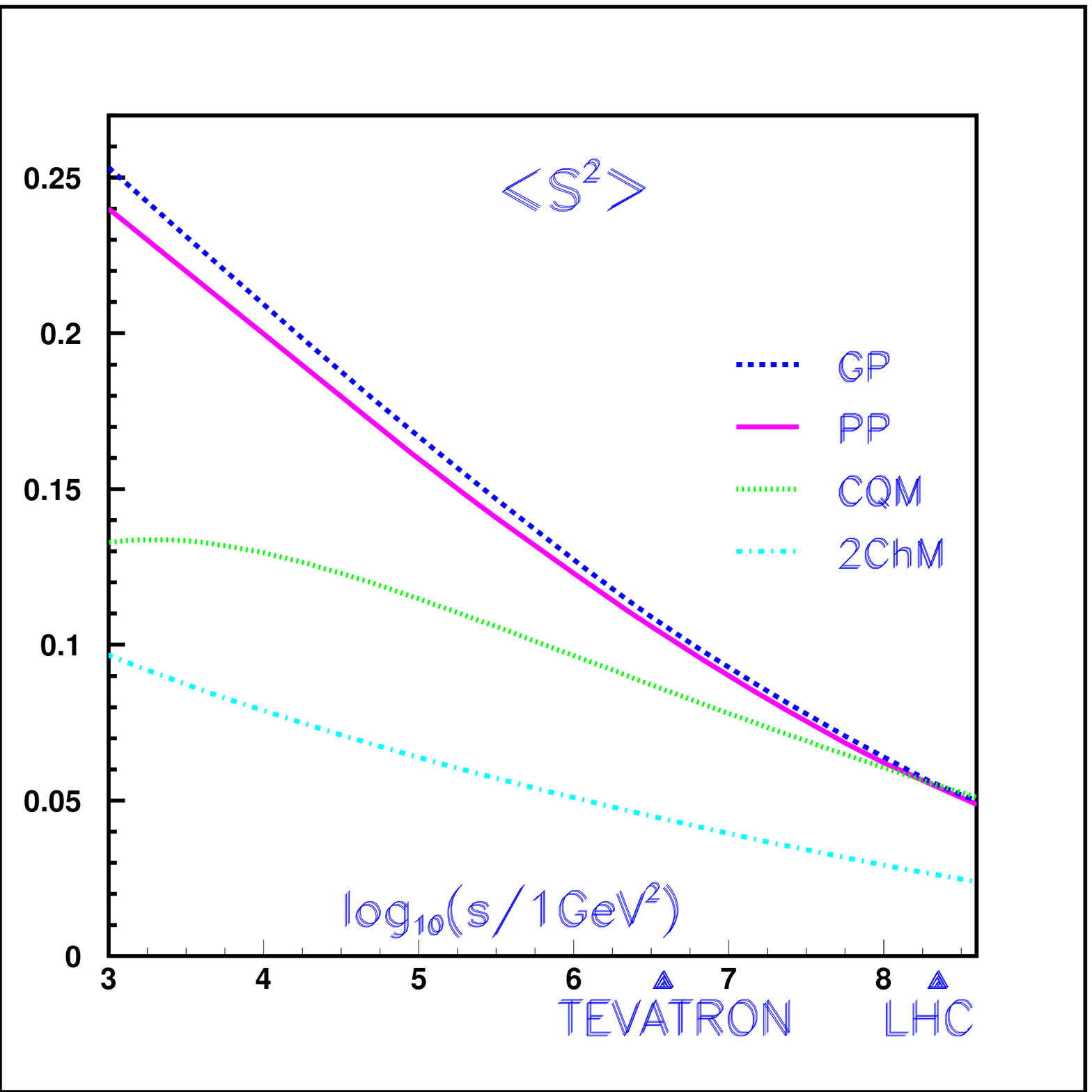,width=160mm}}
\caption{\it Energy dependence of the survival probability for
exclusive central di-jet production as calculated with the four models
we have considered.}
\label{sp}
}

In Section II.C we have explored the two component structure of
$\frac{d\sigma_H}{d^2b}$, (Eq.\eq{DSB2}).
\fig{sp2r} examines the impact of the inelastic component in the
calculation of SP for our process in a single channel model with a PP
profile. The results show a relatively small difference
(within 10\% accuracy) between a single
elastic component and the sum of elastic and non elastic components
(see II.C for details).
An increase of the fraction of the
inelastic production does not change the value of the SP significantly.

\FIGURE[ht]{
\centerline{\epsfig{file=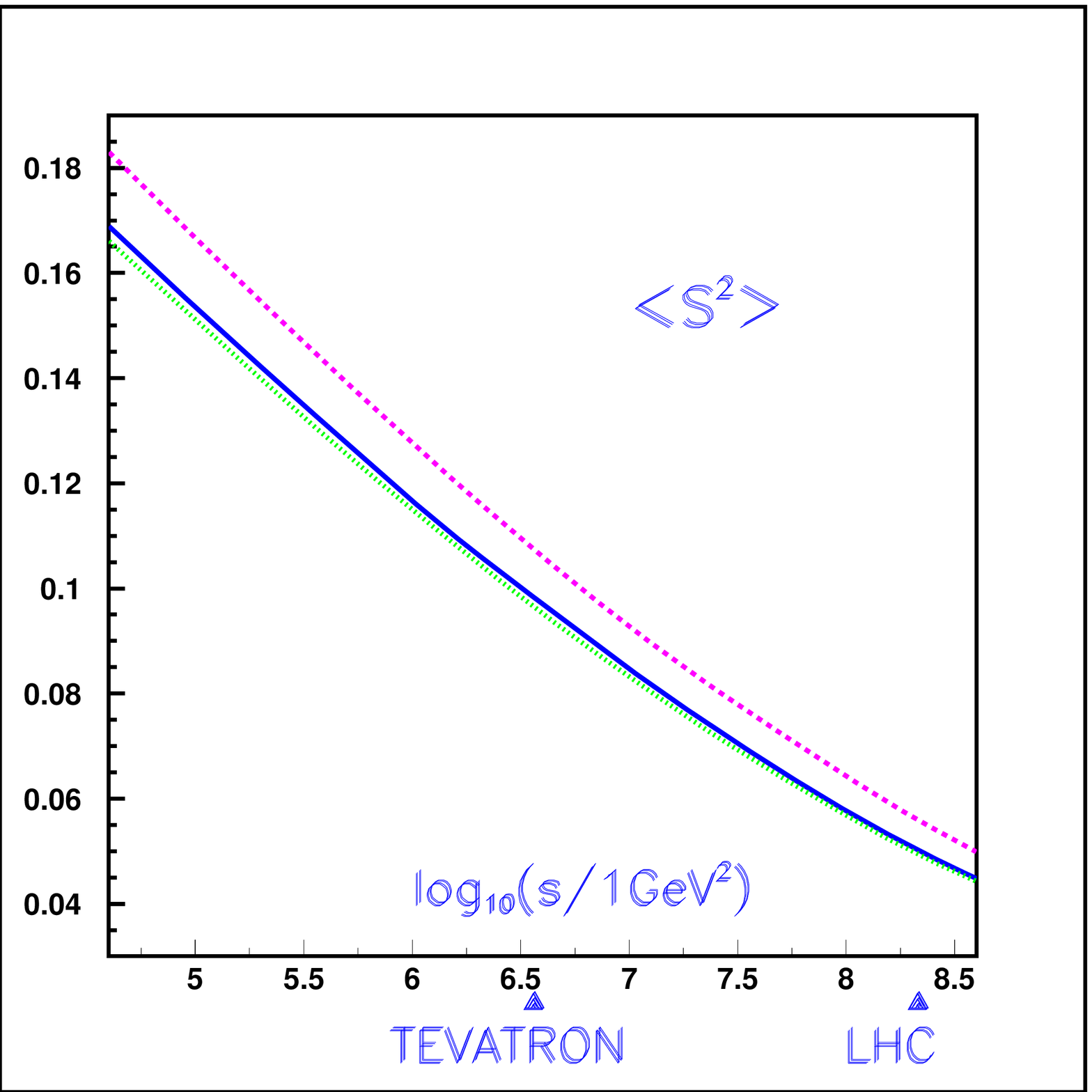,width=100mm}}
\caption{\it Energy dependence of the survival probability for di-jet
production in the eikonal single channel model with
a PP profile. The upper
curve corresponds to the final state production of two nucleons.
The lower curves
take into account the inelastic production of excited states
assuming either \eq{DSB2}, or
that at t=0 the elastic and inelastic
production have the same amplitude. The two last cases can hardly be
distinguished.}
\label{sp2r}}

\fig{sp} details the SP predictions of the four models we have
examined.
We aim, on the basis of these
predictions, to suggest
upper and lower bounds for the SP corresponding to
exclusive central diffractive di-jet production at the LHC.
Examining \fig{sp} we observe that SP calculated in the two channel
model is consistently lower than the corresponding single channel model
and CQM predicted SP values. This may serve as a guide in our attempt to
suggest a margin of error in the determination of SP at the LHC.

When assessing to higher SP bound,
we note that the CQM SP is much lower than the
single channel values at $W\,<\,1000\,GeV$,
with a difference that gets smaller with increased
energy. The two predictions are approximately the same, 6.0$\pm 0.1\%$,
at the LHC energy and they cross just above the LHC.
This suggests a 5-6$\%$ as the upper bound for the calculated SP at the
LHC.
We consider the SP estimates with models neglecting the
diffractive channel rescatterings to over-estimate the
calculated SP output.
Prudency suggests, thus, an upper SP bound of 4- 5$\%$,
which is
moderately smaller than the predictions of the single channel models
and CQM.

The two channel eikonal prediction for exclusive central diffractive
di-jet production at the LHC is 2.7$\%$, compared with
3.6$\%$ for the corresponding inclusive central di-jet
production\cite{heralhc}. The two channel input is
$v\,=\,\sqrt{3}$, $B_{GG}\,=\,3.6 GeV^{-2}$ and
$B_{GG}^{in}\,=\,1.0 GeV^{-2}$.
Our two channel predicted SP at the LHC is almost identical to the
KKMR\cite{KMRSD} value. This is very supportive
of a SP = 2.5-3.0$\%$ at the LHC.

It is instructive to compare our model with the KKMR model\cite{KMRSD}.
The two models are defined as "two channels", but are, actually,
rather different.
\newline
1) Our two channel eikonal definition, for either soft or hard
diffraction, consistently refers to
two possible modes of soft rescattering, i.e. elastic and diffractive.
Accordingly, both elastic and
diffractive states in a $p-p$ scattering are presented as a linear
combination of our two orthogonal base wave functions
$\Psi_1$ and $\Psi_2$.
In our two channel model we neglect the
double diffraction channel, which is exceedingly small in the ISR-Tevatron
energy range. The screening
opacities of our input eikonal matrix
have different b=0 normalizations, i.e. different $\nu$ values,
and different b-dependences which reflect our
different input for the elastic and diffractive forward differential
slopes.
\newline
2) In the KKMR model for soft interactions the eikonal matrix is defined
in a similar way to ours. Note, though, that their diffractive eikonal
components are restricted to low diffractive mass.
The screening opacities have
the same b-dependence,
identical to the b-dependence of the single channel elastic
opacity, having a different b=0 normalizations.
Unlike our input, KKMR do not neglect the double diffractive channel.
\newline
3) It is no surprise that KKMR obtain a very high LHC prediction for
$\sigma_{dd}=9.5mb$, which is comparable to their predicted $\sigma_{sd}$.
KKMR predict, thus, an inelastic diffractive cross section
$\sigma_{sd}\,+\,\sigma_{dd}\,=\,18.9-24.9\,mb$, relative
to a $\sigma_{tot}\,=\,99.1-104.5\,mb$. Our predction is
$\sigma_{sd}\,=\,12.0\,mb$, which may be corrected by a small value
$\sigma_{dd}$, relative to a $\sigma_{tot}\,=\,103.8\,mb$.
Note that we were not successful in reproducing the KKMR fit
for $\sigma_{el}$ and $B_{el}$.
KKMR have not published a detailed prediction for $\sigma_{sd}$.
Their quoted LHC value for single diffraction is comparable to ours.
\newline
4) In the KKMR model the hard process has two dynamical components treated as
two eikonal chennels.
Two sets, with two components each, were studied
and have been shown to produce very similar results.
In our two channel model the hard diffractive proces has two components
associated with the possibility of either an elastic or inelastic
diffractive initial rescattering preceeding the hard process. The SP is
calculated accordingly.
\newline
5) In our model
we make a weak assumption in which the ratio between the elastic
and diffractive couplings is the same for soft and hard Pomerons.
Consequently,  we are able to associate
the hard amplitude with HERA DIS experimental data resulting in eikonal
opacities which are different in their b=0 normalization as well
as their b-dependence.
KKMR make a much stronger assumption in which all opacities, soft and
hard,
have the same b-dependence and a normalization which is determind by the
above coupling ratio assumption.
\newline
6) Regardless of the above the actual KKMR SP results are remarkably close
to ours. This result deserves a clarification in the future.

\section{Transverse
momentum dependence of the cross section}
In the following investigation we have calculated the dependence of
the output di-jet cross section on $p_{1,t}$ and $p_{2,t}$. This
calculation provides the differential information on the t
dependence of the cross section and SP of the process. The
importance of this differential calculation is that it offers a
refined method to discriminate between different models and/or
parametrizations leading to compatible values of SP.
 We define a damping factor
\beq \label{D}
D_{SP} \Lb
p_{1,t},p_{2,t} \Rb\,\,=\,\,\frac{
\frac{d \sigma}{d^2 p_{1t}\,d^2\, p_{2t}\, d y_1\,d y_2} }
{ \int d^2 p_{1t} \,d^2\, p_{2t},
\frac{d \sigma }{d^2
p_{1t}\,d^2\, p_{2t}, d y_1\,d y_2} }.
\eeq
In a single channel eikonal model,
\beq \label{D1}
D_{1C}\,=\,\frac{|\int\,d^2\,b_1\,d^2
b_2\,e^{i\,\vec{p}_{1t}\,\cdot\,\vec{b}_1\,+\,i\,\vec{p}_{2t}\,
\cdot\,\vec{b}_2}
A_H(b_1)\,A_H(b_2)\,\,A_S \Lb (\vec{b}_1\,+\,\vec{b}_2 )^2 \Rb|^2}
{\int\,d^2\,b_1\,d^2
b_2 \,| A_S\Lb (\vec{b}_1\,+\,\vec{b}_2 )^2 \Rb
A_H(b_1)\,A_H(b_2)|^2}.
\eeq

From its definition the value of the SP
is dependent on the b-distribution of the soft amplitude profile.
Should we have an experimental information on the dependence of
$D_{SP}$ on $p_{1,t}$ and $p_{2,t}$, we hope to be able
to invert the above procedure and obtain information on
the form of the b-profile from the differential properties of the SP.

For the generalization to a two channel model we
insert Eqs. \ref{2C12} - \ref{2CT4} into the expression
for the two channel SP.
Using the amplitude, given by \eq{2C12},
we can carry out the integration over $d^2 b_1$ analytically,
obtaining a much simpler expression.
\beq \label{D2}
D_{2C}\,=\,\frac{|A|^2}{\int d^2 p_{1,t}\, d^2 p_{2,t} |A|^2},
\eeq
where $ A = A_1 \,+\,A_2\,+\,A_3$. Defining $\vec{b}\,
=\,\vec{b}_1 \,+\,\vec{b}_2$
and $\vec{p}_{12}\,=\,\vec{p}_{1,t}\,-\,\vec{p}_{2,t}$,
we obtain
\be
A_1 \,\,&=&\,\,\frac{1}{4 \pi^2 B^2_{el}}\,\int\,d^2 b\,
e^{ - \frac{\Omega(b)}{2}}\,s(b)\,\exp \left(
- \frac{b^2}{4 B^2_{el}}\,-\,\frac{1}{4} B_{el}\,p^2_{12}
\right)\,J_0\left(\frac{b}{2}|\vec{p}_{1,t}\,+\,\vec{p}_{2,t}|\right),
 \label{ASH1}\\
A_2 \,\,&=&\,\,-\frac{2 \alpha\,\beta}{4 \pi^2 B_{el}\,
B^2_{inel}}\,\int\,d^2 b\,\,e^{ -
\frac{\Omega(b)}{2}}\,\,s_D(b)\,  \label{ASH2} \\
&\cdot&\exp \left(- \frac{1}{2 (B_{inel} \,+\,B_{el})}\,\{ b^2
\,+\,B_{inel}\,B_{el}\,p^2_{12}\}\right)\,\,J_0\left(\frac{b}{B_{inel}
\,+\,B_{el}}\,| B_{el} \,\vec{p}_{1,t}\,+\,B_{inel}\,\vec{p}_{2,t}|
\right), \nonumber\\
A_3 &=& A_2 \left( \vec{p}_{1,t} \leftrightarrow \vec{p}_{2,t} \right).
\label{ASH3}
\ee

\fig{df} and \fig{df1} show the single channel di-jet factor $D_{1C}$
with GP and PP b-profiles as a function of the transverse momenta of the
produced protons in the final state.
We predict a typical minima
at $p_{1,t} \,\approx 1.75 - 2 \,\gev$.  This  behaviour is a
manifestation of the wave nature of our diffractive scattering
process. It depends on the scale of the soft profile b-distribution.
From \fig{df} and
\fig{df1} one can see that the minima $p_t$ position
depends on both the energy of $p_{1,t}$ and $p_{2,t}$ and the angle
$\theta$ between them.
As stated, we hope that a future measurement of this
dependence will provide valuable information which will add to our
knowledge of the soft interaction amplitude in QCD.
To get some initial information on the angular distribution,
we have chosen (for convenience)
$\theta\,=\,0,\,\pi/2, and\,\pi$.

In \fig{ds} and \fig{ds1} we compare the transverse momentum distribution
of the di-jet production differential cross section with the corresponding
elastic cross section. These figures illustrate that the di-jet
cross section has quite a different structure of minima than the
elastic cross section, where the positions of these minima move to
smaller values of $p_t$ at higher energies. Comparing with the
experimental data (for data see Ref.\cite{BAPR} and
references therein), we deduce that  this data is compatible
with the PP profile describing the structure of the
minima and maxima in proton-proton collisions in the energy range of
$W = \sqrt{s} = 20 - 65 \,GeV$.
Our parametrizations for the profile
function give a range of predictions for the di-jet production at
the LHC. The difference between the GP and PP parametrizations, as far as
$p_t$ dependence of the di-jet cross section is concerned, is much
smaller than for the elastic cross section.

A most interesting result is the striking difference between
the t-dependence of the damping factors,
defined in \eq{D}, obtained in the single and two channel eikonal models.
This difference is clearly seen in \fig{2ds} in which GP profiles were
used.
The single channel model leads to a dip and to a slope at
$p_{1,t}$=0 which is equal to $B_{GP}$ without a suppression due to SP.
In the two channel eikonal model we do not
have any dip at $p_{1,t} \,\leq 2\,\,GeV$ and the slope
turns out to be much smaller than the slope
obtained from the single channel eikonal model.

This result can be anticipated from the general formula of
\eq{2C9} and \eq{2C11} which show that
the damping factor is proportional to a
contribution of the order of $\beta$ while the elastic cross
section is proportional to $\beta^2$
(see Ref.\cite{GLMLRG}).

\FIGURE[ht]{
\begin{tabular}{c c }
\includegraphics[width=0.5\textwidth]{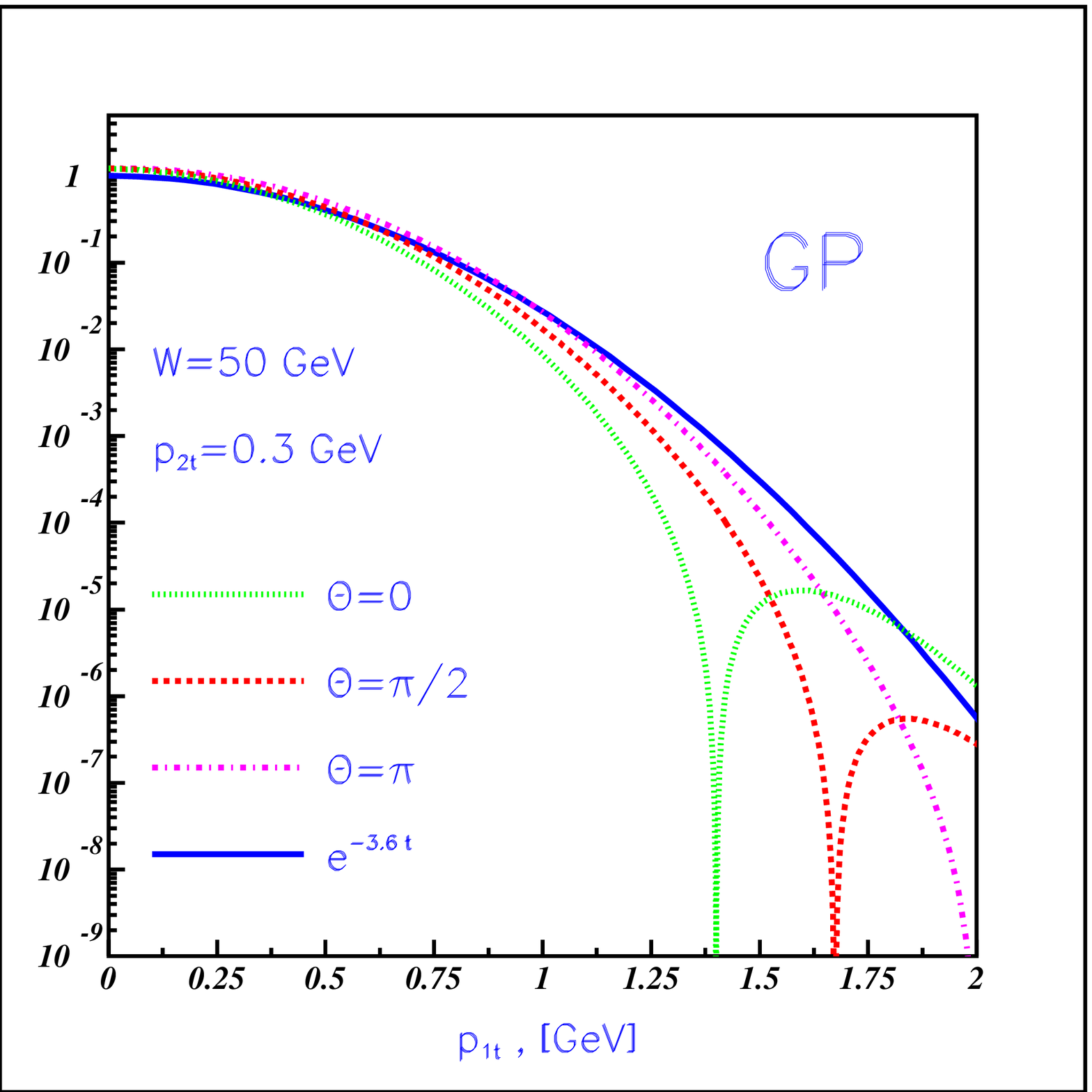} &
        \includegraphics[width=0.5\textwidth]{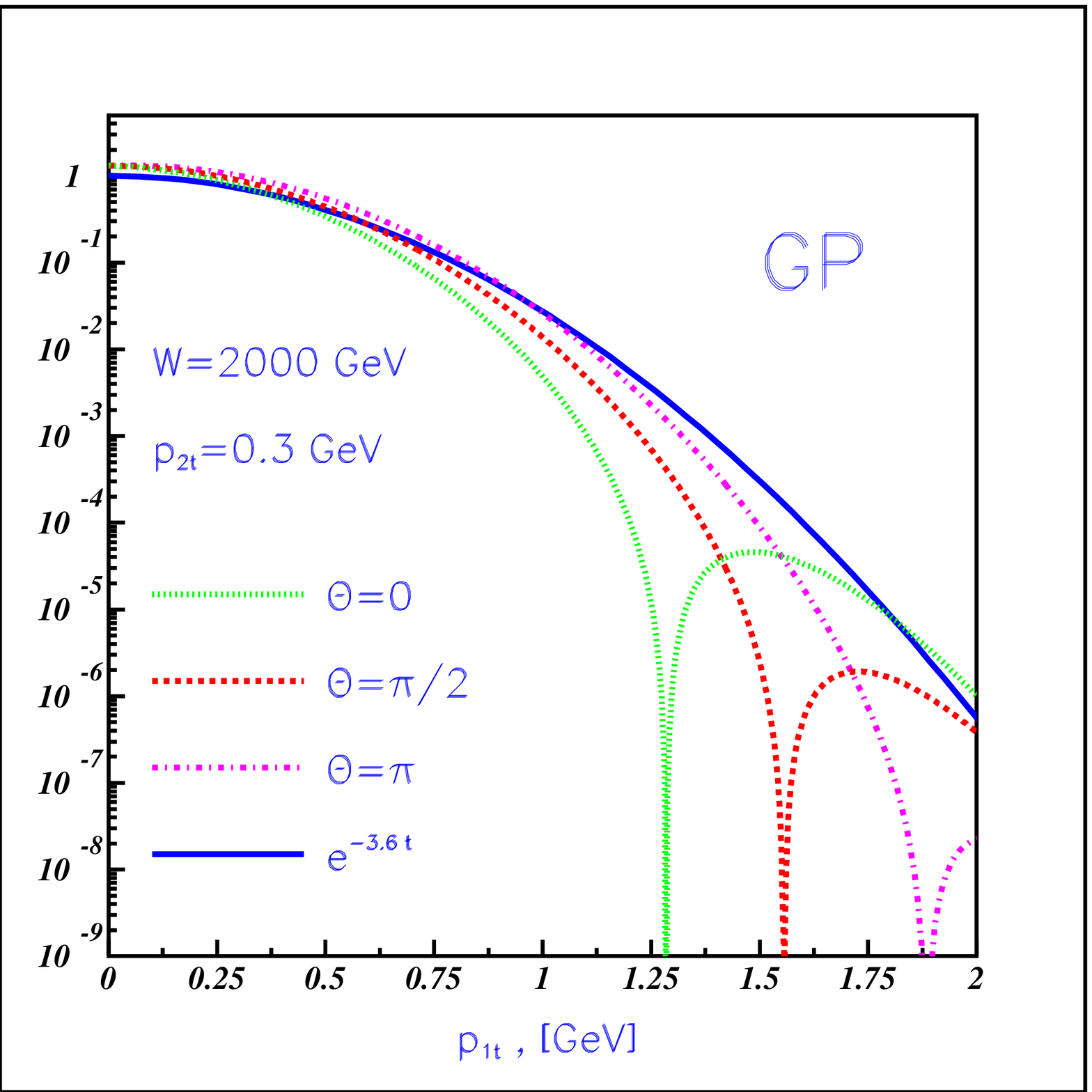}\\
        \\\fig{df}-a & \fig{df}-b\\
        \includegraphics[width=0.5\textwidth]{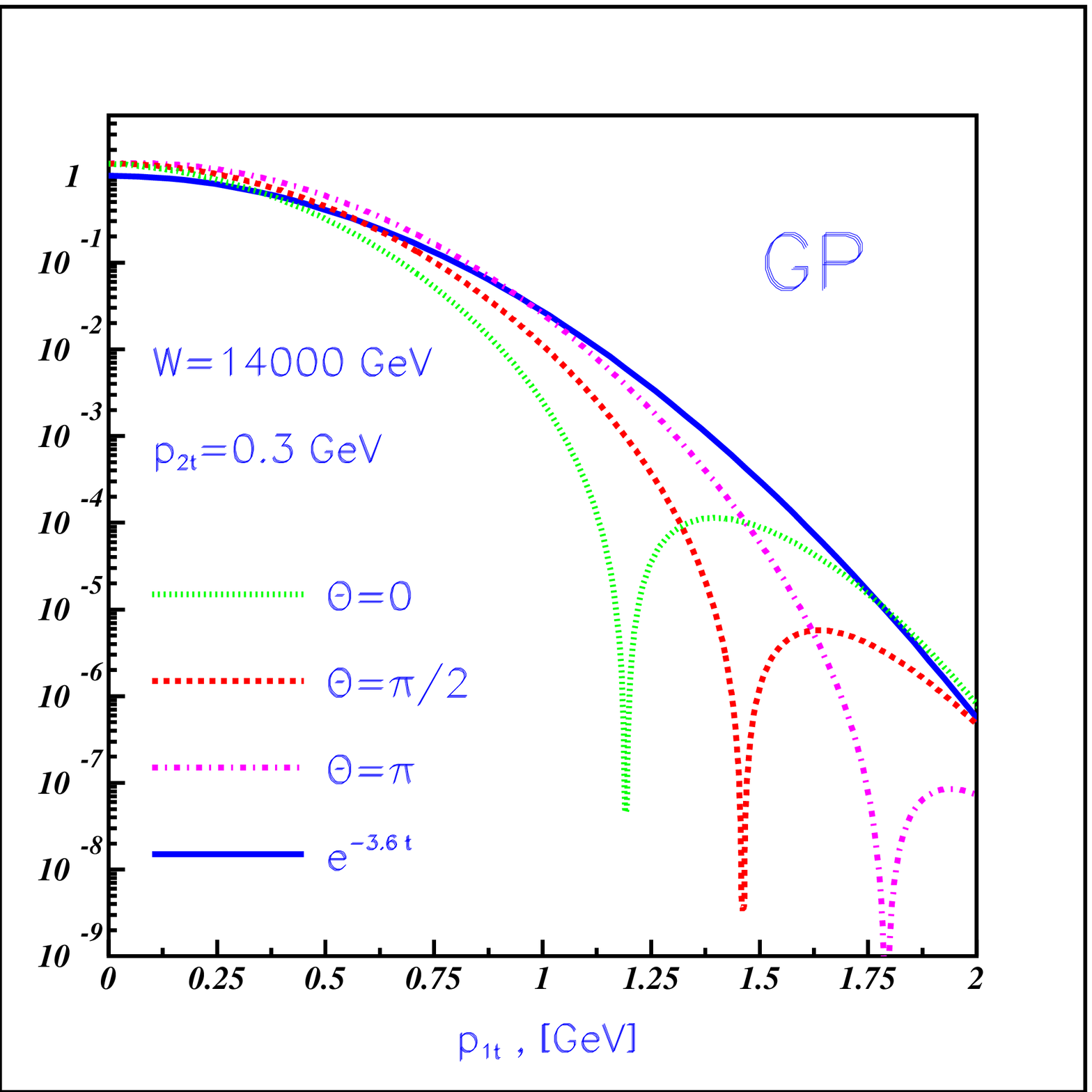} & \\
        \fig{df}-c &\\
\end{tabular}
\caption{\it Transverse momentum dependence of the cross section
factor $D$ in a single channel model with a Gaussian b-profile (GP)
for various $\theta$ values.
We also show an $\exp(- 3.6 |t|)$ dependence corresponding to to the
hard slope $B_{GG}$.}
\label{df}
}
\FIGURE[ht]{
\begin{tabular}{c c }
  \includegraphics[width=0.5\textwidth]{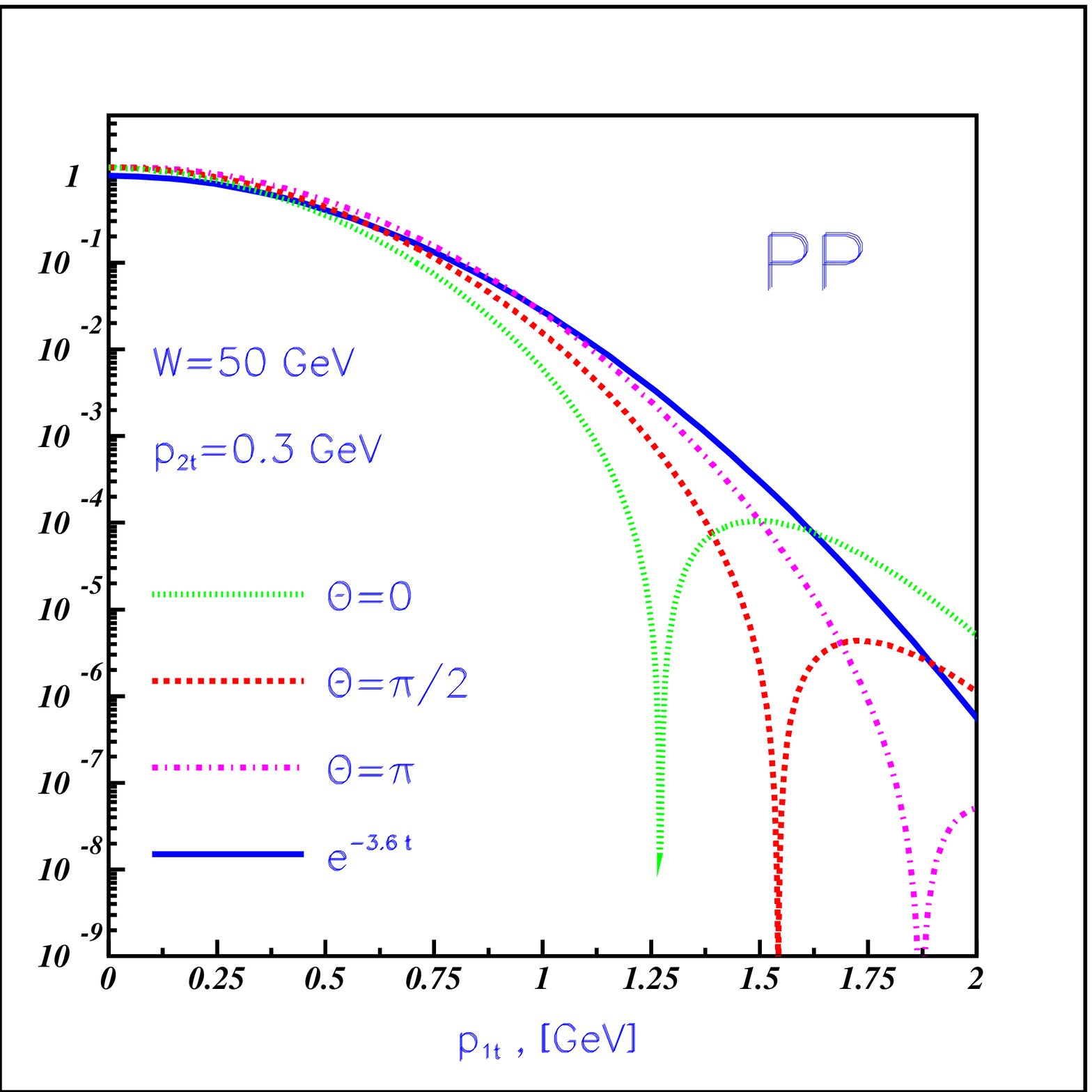} &
 \includegraphics[width=0.5\textwidth]{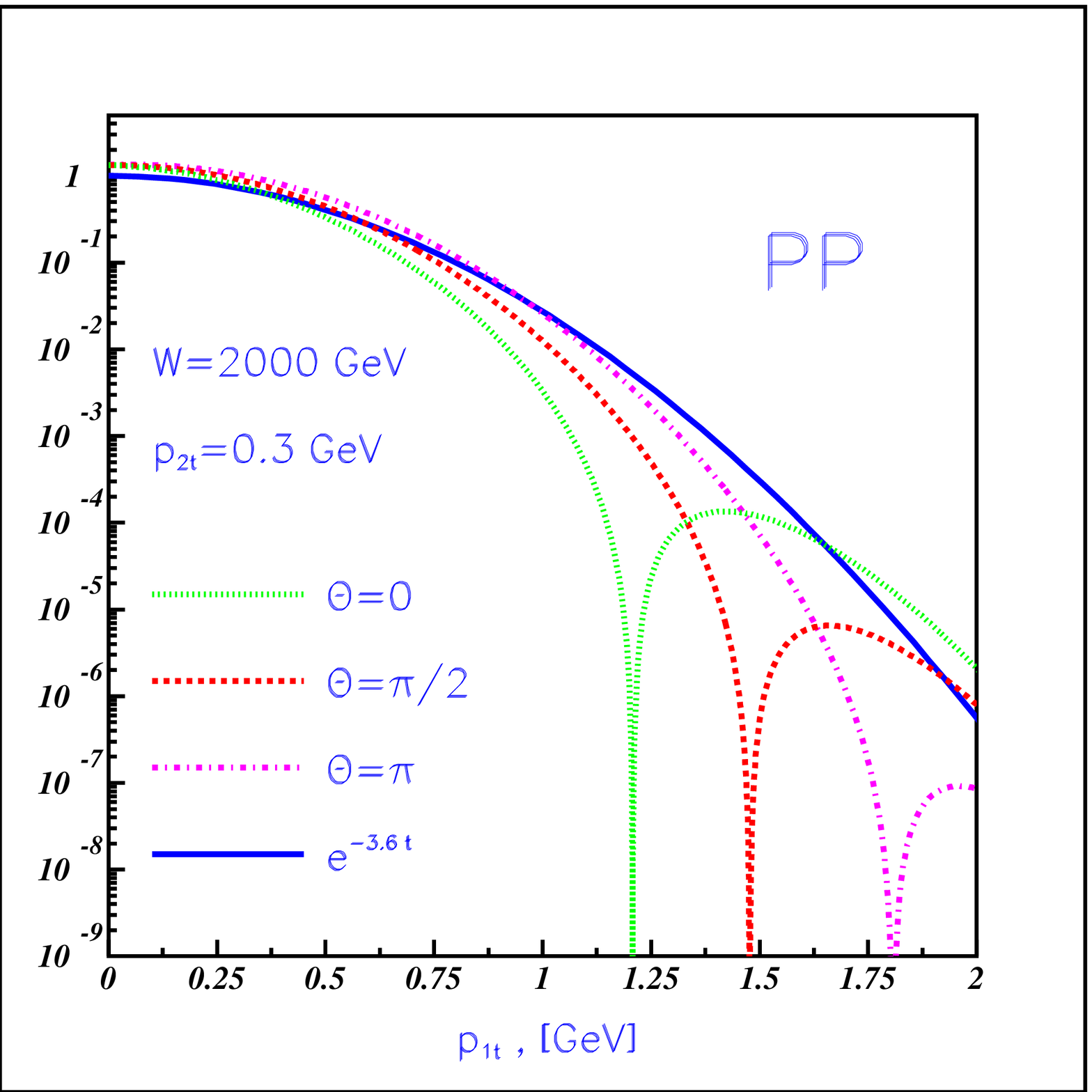}  \\
 \fig{df1}-a  & \fig{df1}-b \\
\includegraphics[width=0.5\textwidth]{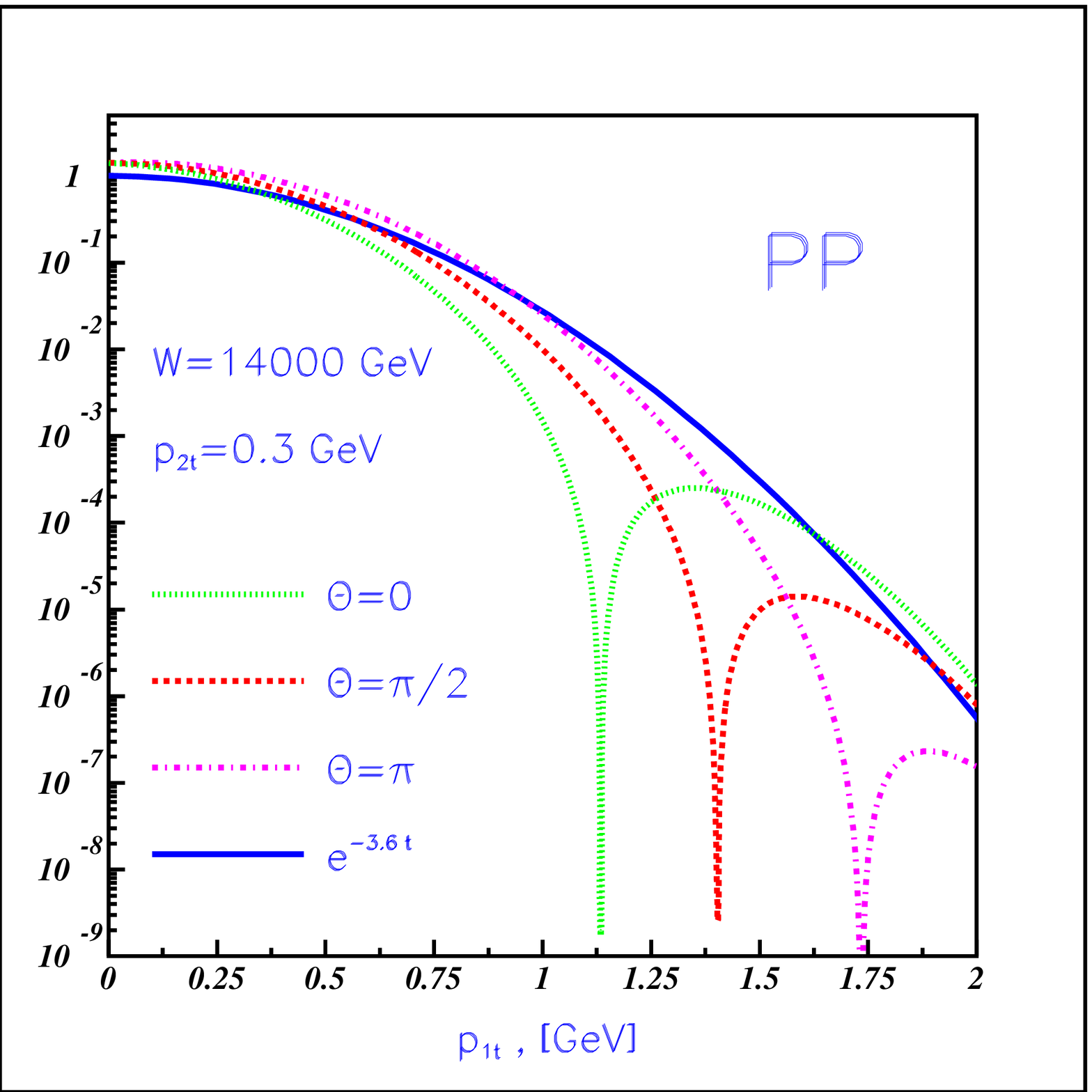}\\  \fig{df1}-c\\
\end{tabular}
\caption{\it Transverse momentum dependence of the cross section
factor $D$ in a single channel model with a power like b-profile (PP)
for various $\theta$ values.
We also show an $\exp(- 3.6 |t|)$ dependence corresponding to to the
hard slope $B_{GG}$.}
\label{df1}
}

\FIGURE[ht]{
\begin{tabular}{l l }
        \includegraphics[width=0.5\textwidth]{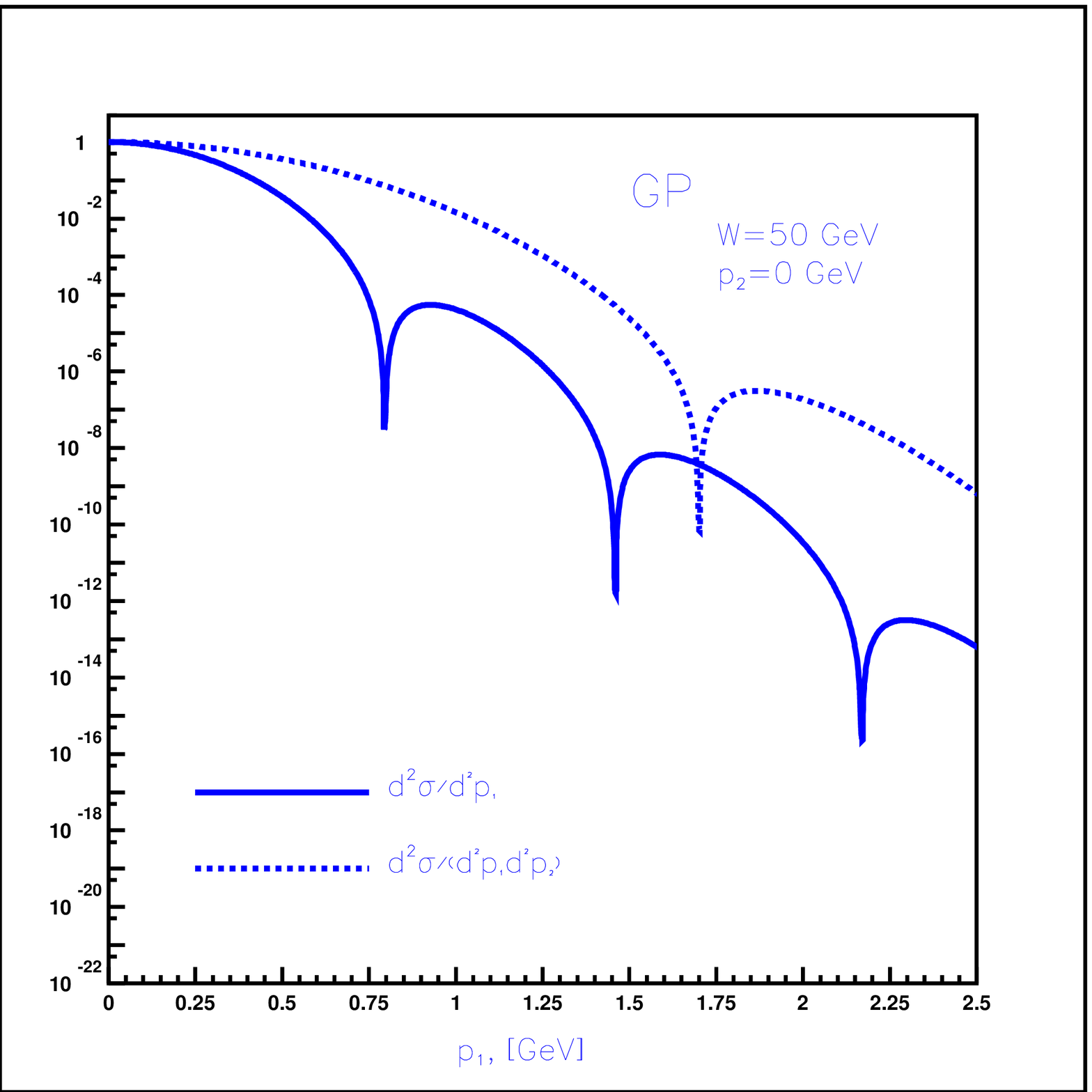} &
 \includegraphics[width=0.5\textwidth]{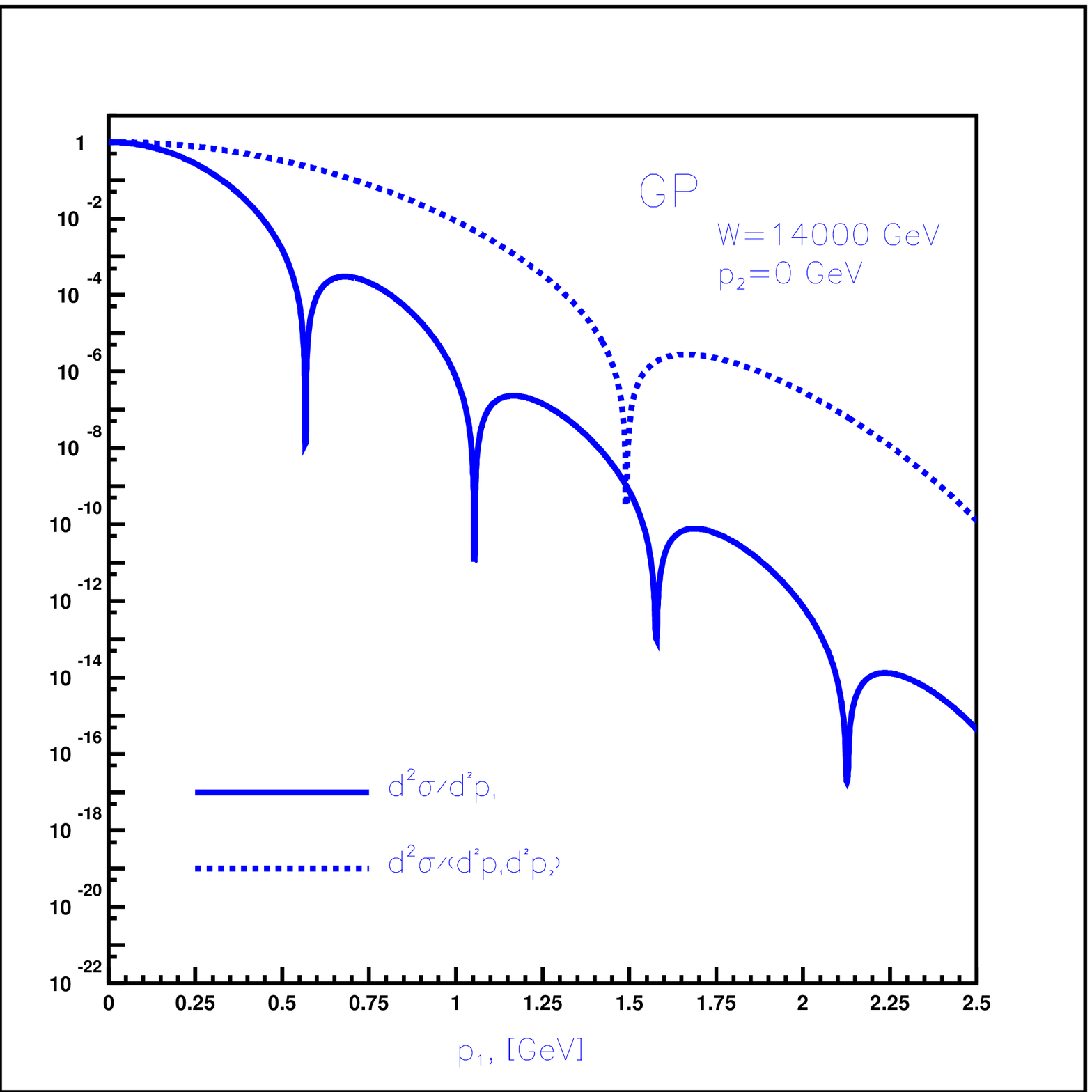}
 \\
\fig{ds}-a  &
  \fig{ds}-b
\\
\end{tabular}
\caption{\it Transverse momentum dependence of the cross section
factor $D$ in a single channel model with a Gaussian b-profile (GP)
compared with the corresponding elastic cross section.}
\label{ds}
}
\FIGURE[ht]{
\begin{tabular}{l l }
 \includegraphics[width=0.5\textwidth]{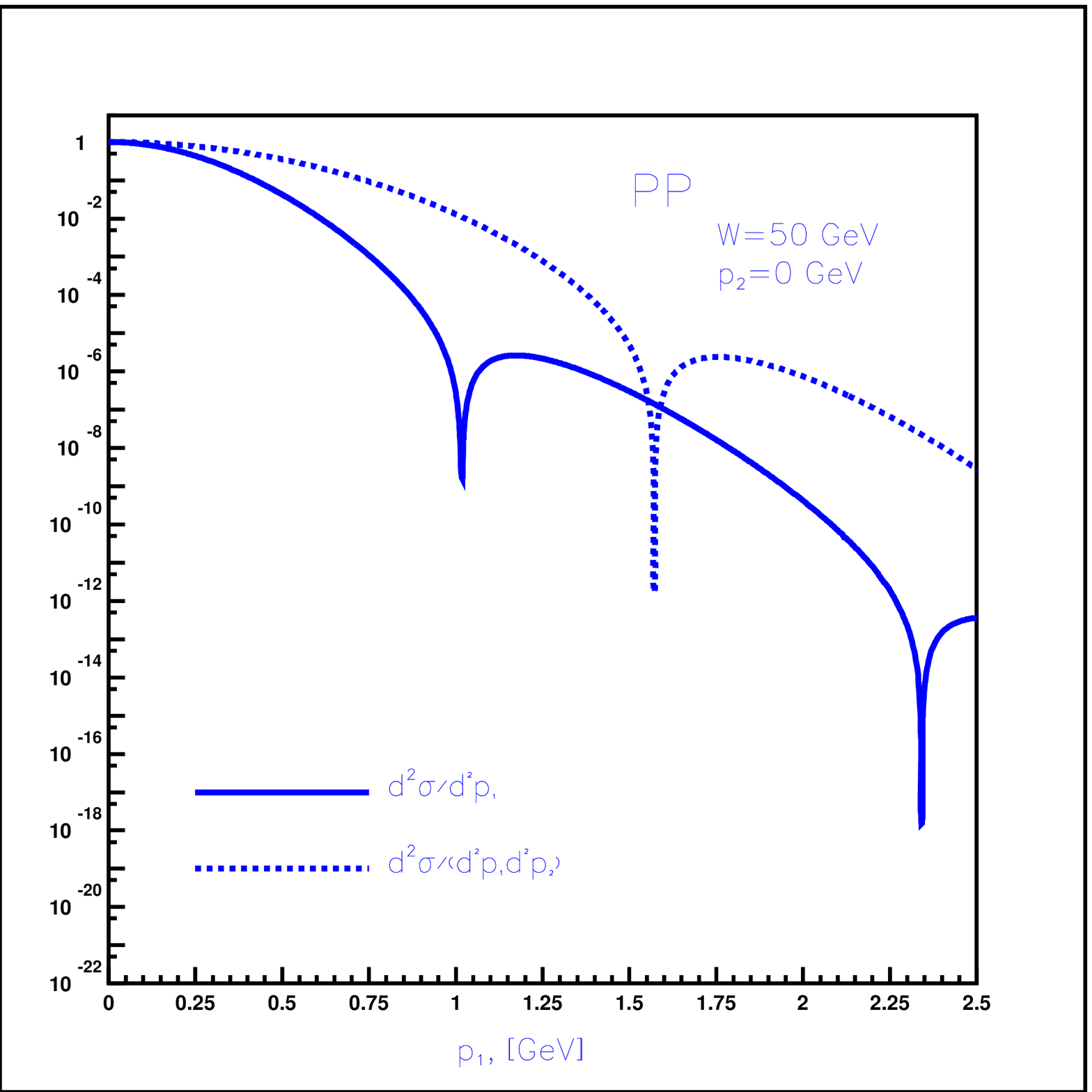} &
  \includegraphics[width=0.5\textwidth]{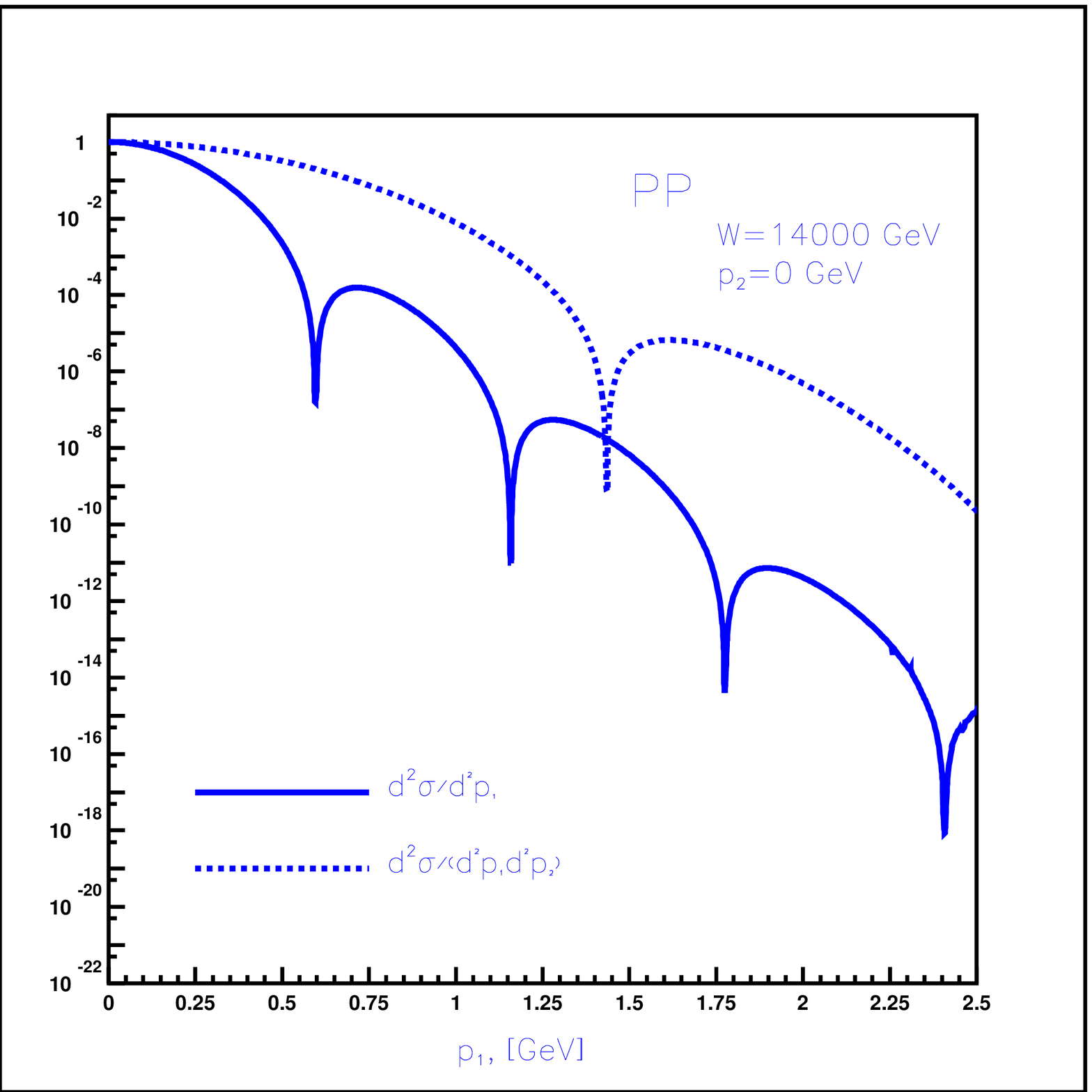}\\
    \\ \fig{ds1}-a  & \fig{ds1}-b\\
\end{tabular}
\caption{\it Transverse momentum dependence of the cross section
factor $D$ in a single channel model with a power like b-profile (PP)
compared with the corresponding elastic cross section.}
\label{ds1}
}

\FIGURE[ht]{
\begin{tabular}{c c }
       \includegraphics[width=0.5\textwidth]{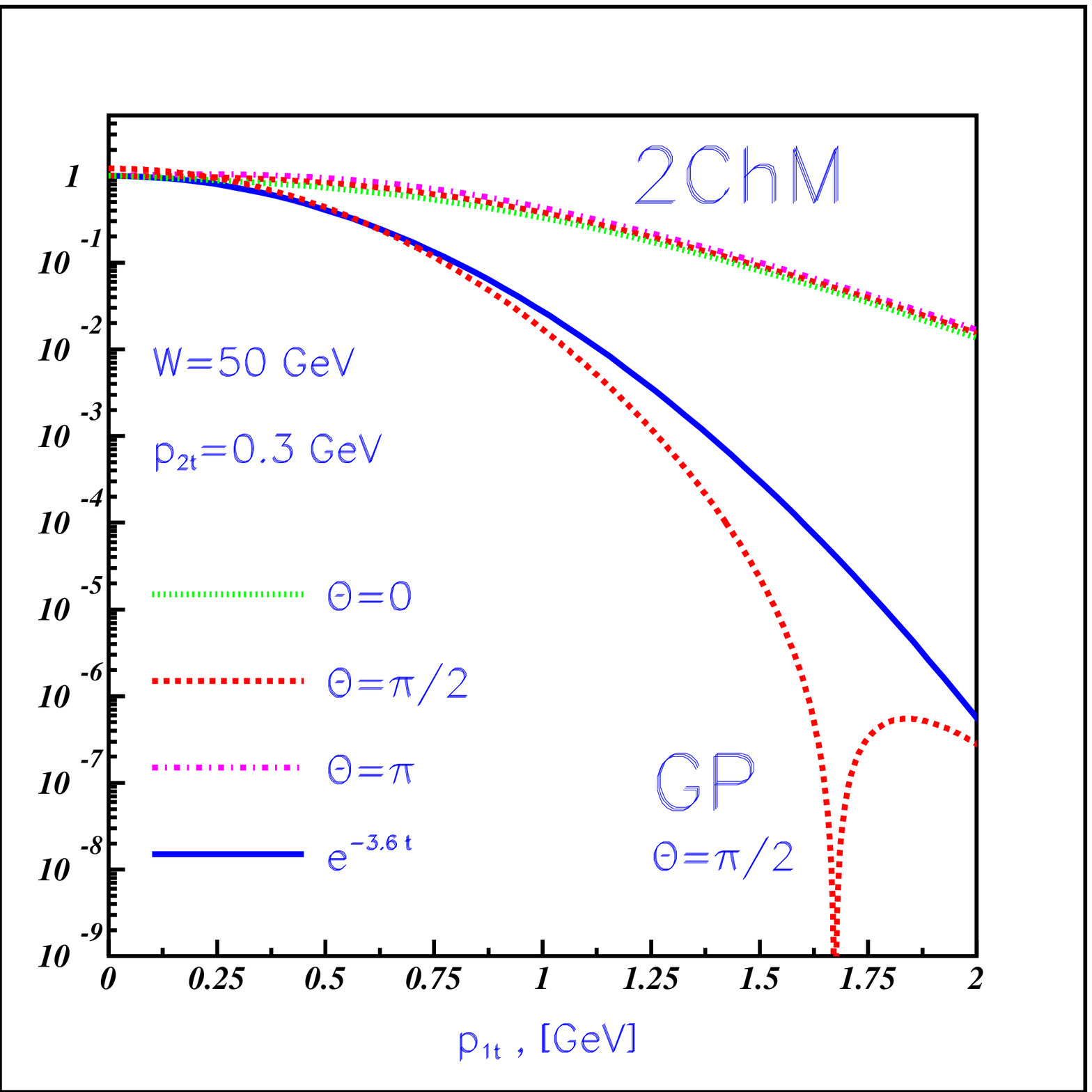} &
\includegraphics[width=0.5\textwidth]{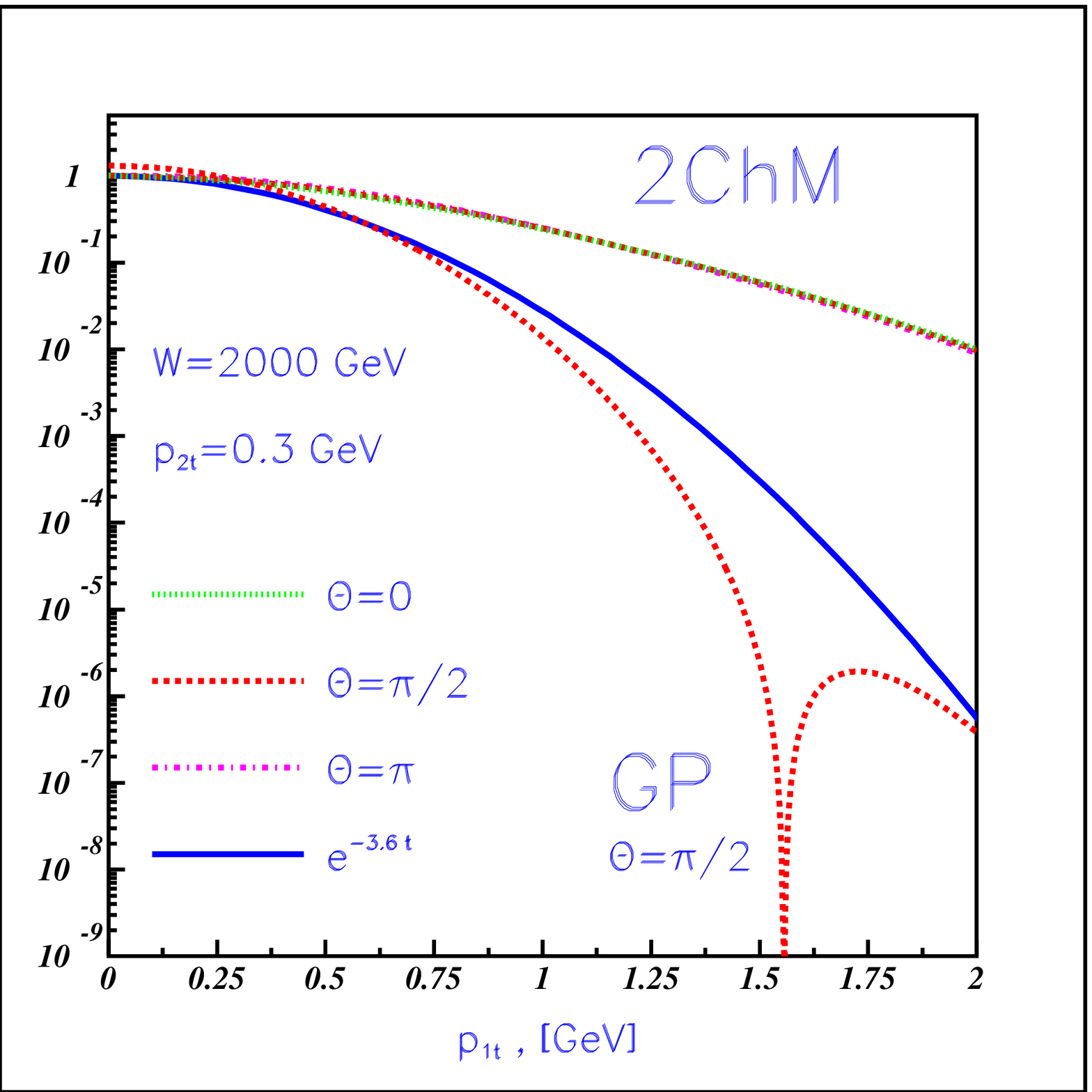}\\
   \fig{2ds}-a  & \fig{2ds}-b \\
\includegraphics[width=0.5\textwidth]{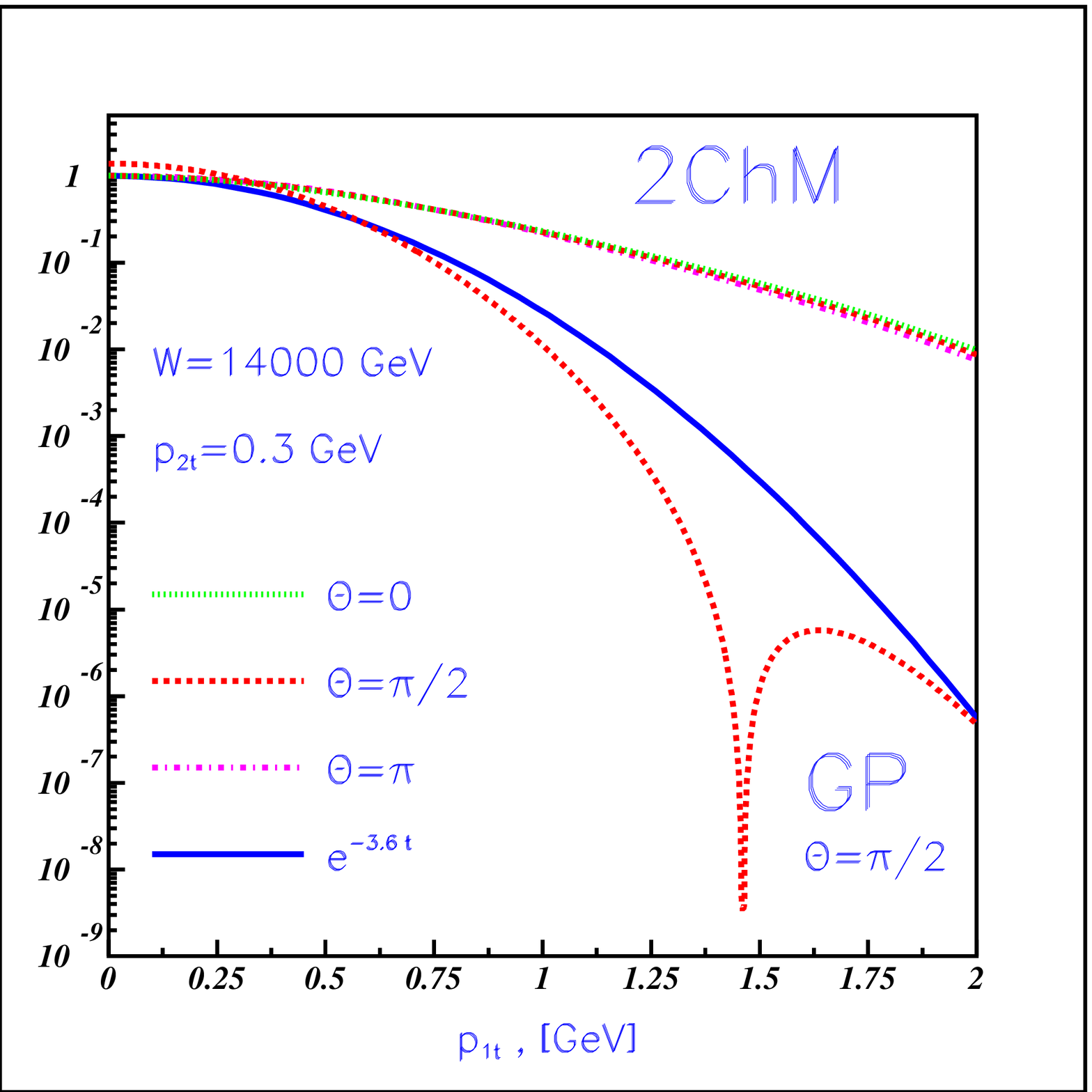}\\
\fig{2ds}-c\\
\end{tabular}
\caption{\it Transverse momentum dependence of the cross section
factor $D$ in a two channel model with Gaussian b-profiles (GP)
for various $\Theta$ values,
compared with the corresponding single channel elastic cross section
with $\theta=\pi/2$.
We also show an $\exp(- 3.6 |t|)$ dependence corresponding to to the
two channel input hard slope $B_{GG}$.}
\label{2ds}
}

\section{Conclusions}
In this paper we have calculated the large rapidity gap survival
probability $< |S|^{2} >$, for exclusive central dijet production at
the LHC. Our assessments are of particular interest as the SP we
have calculated correspond also to a central exclusive diffractive
production of Higgs. The unique feature of our treatment is that we
explicitly included the impact parameter dependence of the hard
amplitude in our evaluation of SP. This b-dependence appears in the
proton-hard Pomeron vertex, whose structure was deduced from
information obtained from the DIS production of $J/ \psi$.

The calculation of the LRG SP requires knowledge
of both the hard
and soft components of the hadron-hadron interactions.
As there is no
reliable theory for the soft component, we employed four
phenomenological eikonal models to describe the soft processes.
These models are:
\newline
1) A single channel model with an exponential dependence in t for the
proton-soft Pomeron vertex.
\newline
2) A single channel model with a power-like behaviour for the proton-soft
Pomeron vertex.
\newline
3) The CQM, which is a non Regge two component soft model, where we have
assumed that this vertex is given by
the proton electromagnetic form factor approximated by
a dipole.
\newline
For cases 2) and 3), we have to integrate
over b numerically, while for case 1), we have an analytical expression.
\newline
4) To these we add the two channel model whose
main advantage is that it is
the only model considered that properly reproduces the available
diffractive data.

All four models considered were utilized to calculate the SP of
an exclusive central hard di-jet production. The two channel model
predicts for this SP a value of 0.027, which is almost identical to
the KKMR prediction. These should be compared with a 0.036 value for
the equivalent inclusive calculation\cite{heralhc}.
The other three models have compatible prediction of
0.06. We consider this value to be some what
over estimated as these models neglect the diffractive rescattering which
increases the eikonal overall screening and, thus, reduces the SP.
We prudently conclude that the LHC SP we have studied
lies between 0.03 -  0.05.

The single channel cross-section for the dijet
production as a function of $p_{t}$, has a typical
structure of minima at certain values of $p_{t}$, which is
dependent both on the angle between $p_{t1}$
and $p_{t2}$ of the two
jets, and on their energy. Positions of the
minima are model dependent.

The most interesting result of this paper is the fact that
the two channel model predicts quite
different dependence of the damping factor
versus the transverse momenta.  The central prediction is that
there are no dips for
$p_t \,\leq \,2\,GeV$. Also, the slope at $p_t=0$ turns out
to be much smaller than for the one channel eikonal model.
Both features stem from the general properties of the two channel model
and we firmly believe that this
prediction should assist to select the
appropriate models for the description of SP
directly from the experimental data. Since the general behaviour of the
damping factor depends weakly on
energy we suggest assessing the damping factor at the Tevatron
energies.
We would like to stress that the calculation of the damping factor show
such simplicity and elegance that we believe they have a deeper meaning
than being just a consequence of a particular model.

\section*{Acknowledgements}
We would like to thank  S. Bondarenko, V. Khoze and  E. Naftali
for useful discussions on the subject of this paper.
This research was supported in part  by the Israel Science Foundation,
founded by the Israeli Academy of Science and Humanities
and by BSF grant \# 20004019.
The work of A. Prygarin was supported by Minerva Fellowship Foundation,
Max-Planck-Gesellschaft.
\appendix



\begin{thebibliography}{99}

\bibitem{DOK}
 Yu. L. Dokshitzer, V. Khoze and S.I. Troyan, Proc.{\it ``Physics in
Collisions 6"}, p. 417,  ed. M. Derrick, WS 1987;   Sov. J. Nucl.
Phys. {\bf 46},  712 (1987).\\
Y.~L.~Dokshitzer, V.~A.~Khoze and T.~Sjostrand,
  Phys.\ Lett.\  {\bf B274}, 116 (1992).


\bibitem{BJ}
 J. D. Bjorken,  Int.\ J.\ Mod.\ Phys.\  {\bf A7}, 4189 (1992);\,\,\,
 Phys.\ Rev.\  {\bf D47}, 101 (1993).

\bibitem{GLM}
E. ~Gotsman, E.M. ~Levin and U. ~Maor,
Phys.\ Rev.\  {\bf D49}, R4321 (1994).

\bibitem{CZ}
  H.~Chehime and D.~Zeppenfeld,
  Phys.\ Rev.\  {\bf D47}, 3898 (1993).

\bibitem{FLST}
  R.~S.~Fletcher and T.~Stelzer,
  Phys.\ Rev.\  {\bf D48}, 5162 (1993).
\bibitem{GLMSP}
E.~Gotsman, E.~Levin and U.~Maor,
Phys.\ Lett.\ {\bf B309} (1993) 199 (1993);
Phys.\ Lett.\  {\bf B438} 229 (1998). 

\bibitem{GLMLRG}
E.~Gotsman, E.~Levin and U.~Maor,
  Phys.\ Rev.\  {\bf D60} 094011 (1999);
  Phys.\ Lett.\  {\bf B452}, 387 (1999).
\bibitem{DG}
V.~A.~Khoze, A.~D.~Martin and M.~G.~Ryskin,
  Eur.\ Phys.\ J.\  {\bf C24} 581 (2002),
  Eur.\ Phys.\ J.\  {\bf C21} 521 (2001), 
  Eur.\ Phys.\ J.\  {\bf C14} 525 (2000).
\bibitem{DGK}
A.~B.~Kaidalov, V.~A.~Khoze, A.~D.~Martin and M.~G.~Ryskin,
  Eur.\ Phys.\ J.\  {\bf C21} 521 (2001).

\bibitem{BLHL}
  M.~M.~Block and F.~Halzen,
  Phys.\ Rev.\  {\bf D63}, 114004 (2001).

\bibitem{Fermi}
For recent reviews see: K. Goulianos, 
{\it Proceedings of Diffraction 2002, Alushta (Crimea)},
Kluwer Academic Pub. (2002) 13; {\it J. Phys.} {\bf G26} (2000) 716;
{\it Nucl. Phys. Proc. Suppl.} {99A} (2001) 37; hep-ph/0407035.

\bibitem{HERA}
ZEUS Collaboration,
Eur.\ Phys.\ J.\  {\bf C6} 43 (1999);
Phys.\ Rev.\ Lett.\  {\bf 84} 5083 (2000).\\
H1 Collaboration, 
Nucl.\ Phys.\  {\bf B429} 477 (1994);
Phys.\ Lett.\  {\bf B348} 681 (1995);
Zeit.\ Phys.\  {\bf C76}  613 (1997).


\bibitem{LN}
F. Low, Phys. Rev. {\bf D12},  163 (1975).\\ 
S. Nussinov,
Phys.\ Rev.\ Lett.\ {\bf 34},  1286 (1975); 
Phys.\ Rev.\ {\bf D14}, 244 (1976).

\bibitem{KL}
D.~Kharzeev and E.~Levin,
  Nucl.\ Phys.\  {\bf B578} 351 (2000).

\bibitem{KKL}
D.~E.~Kharzeev, Y.~V.~Kovchegov and E.~Levin,
  Nucl.\ Phys.\  {\bf A690} 621 (2001).
\bibitem{SHZ}
 E.~V.~Shuryak and I.~Zahed,
  Phys.\ Rev.\  {\bf D67} YV054025 (2003).\\
M.~A.~Nowak, E.~V.~Shuryak and I.~Zahed, 
  Phys.\ Rev.\  {\bf D64}  034008 (2001)
\bibitem{JANIK}
 R.~A.~Janik,
  Acta Phys.\ Polon.\  {\bf B33} 3615 (2002);
Phys.\ Lett.\  {\bf B500} 118 (2001).
\\
R.~A.~Janik and R.~Peschanski,
  Nucl.\ Phys.\  {\bf B586}, 163 (2000); 
  Nucl.\ Phys.\  {\bf B625}, 279 (2002).
\bibitem{NACHT}
O.~Nachtmann,
  Annals Phys.\  {\bf 209}, 436 (1991).
\bibitem{SU}
F.~Schrempp and A.~Utermann,
  Phys.\ Lett.\  {\bf B543} 197 (2002).
\bibitem{KLT}
D.~Kharzeev, E.~Levin and K.~Tuchin,
  Phys.\ Lett.\  {\bf B547} 21 (2002).
\bibitem{FIIM}
E.~Ferreiro, E.~Iancu, K.~Itakura and L.~McLerran,
  Nucl.\ Phys.\  {\bf A710} 373 (2002).

\bibitem{DL}
A. Donnachie and P.V. Landshoff,  
  Nucl.\ Phys.\ {\bf B244}, 322 (1984);
  Nucl. Phys. {\bf B267}, 690 (1986); 
 Phys. Lett. {\bf B296}, 227 (1992);  
 Z. Phys. {\bf C61}, 139 (1994).


\bibitem{KOTE}
H.~Kowalski and D.~Teaney,
  Phys.\ Rev.\  {\bf D68} 114005 (2003).

\bibitem{heralhc}
E.~Gotsman, E.~Levin, U.~Maor, E.~Naftali and A.~Prygarin,
{\it Proceedings of HERA and LHC Workshop}, CERN Pub. (2005). 

\bibitem{BFKL}
 E. A.~Kuraev, L. N.~Lipatov, and F. S.~Fadin,  
Sov.\ Phys.\ JETP {\bf 45} 199 (1977). \\
Ya. Ya.~Balitsky and L. N.~Lipatov,
Sov.\ J.\ Nucl.\ Phys.\  {\bf 28} 22 (1978).


\bibitem{DGLAP}
 V. N.~Gribov and L. N.~Lipatov, 
 Sov.\ J.\ Nucl.\ Phys.\ {\bf 15}  438 (1972).\\
 G. Altarelli and G. Parisi,  
 Nucl.\ Phys.\ {\bf B126} 298 (1977). \\
 Yu. l. Dokshitser,  
 Sov.\ Phys.\ JETP {\bf 46} 641  (1977). 

\bibitem{DUR2J}
 V. A.~Khoze, A. D.~Martin and M. G.~Ryskin,
  Eur.\ Phys.\ J.\  {\bf C19} 477 (2001) 
  [Erratum-ibid.\  {\bf C20} 599 (2001)];
  Phys.\ Rev.\  {\bf D56} 5867 (1997). 



\bibitem{SOFT}
{\it `` Regge theory of low-$p_t$ hadtronic interaction",} ed. Luca 
Caneschi, 1989,North Holland Pub.

\bibitem{BAPR}
V.~Barone and E.~Predazzi, 
{\it `` High-Energy Particle Diffraction"}, Springer Verlag Pub. (2002). 

\bibitem{BOLE}
S.~Bondarenko and E.~Levin,
{\it ``Proton proton interaction in constituent quarks model at LHC energies,''}
  arXiv:hep-ph/0511124.

\bibitem{GLMTHIS}
E.~Gotsman, E.~Levin, H.~Kowalski, U.~Maor, and A.~Prygarin,
arXiv:hep-ph/0512254 (version 1).


\bibitem{KMRSD}
V.~A. Khoze, A. D.~Martin and M. G.~Ryskin,
  Eur.\ Phys.\ J.\  {\bf C18} 167 (2000). 

\bibitem{expdata}
Particle Data Group, 
{``Review of Particle Physics"}  Eur.\ Phys.\ J.\,{\bf C3} (1998) 1, 
and references therein.


\end{thebibliography}
\end{document}